\documentclass{ar-1col}

\usepackage[numbers]{natbib}
\usepackage{epic,eepic,units}
\usepackage{soul}
\usepackage{psfrag}
\usepackage{latexsym}
\usepackage{longtable}
\usepackage{mathrsfs}
\usepackage{bigstrut}
\usepackage{pifont}
\usepackage{amsmath}
\usepackage{setspace}
\usepackage{subfigure}
\usepackage{mathtools}
\usepackage{mathcomp}
\usepackage{mathdots}
\usepackage{rotating}
\usepackage{array}
\usepackage{multirow}
\usepackage{multicol}
\usepackage{color}
\usepackage{url}
\usepackage{algorithm}
\usepackage{algorithmic}
\usepackage{graphicx}
\usepackage[dvipsnames]{xcolor}
\usepackage{euscript,yfonts,dsfont,bbm,wasysym,pdfsync,arydshln,framed}
\usepackage{amstext}
\usepackage{amssymb}

\usepackage{tikz}
\usetikzlibrary{arrows,shapes,trees,calc,positioning}
\usetikzlibrary{matrix}

\newcommand{\bbR}{\mathbb{R}}
\newcommand{\bbC}{\mathbb{C}}

\newcommand{\eps}{{\epsilon}}

\newcommand{\non}{\nonumber}
\newcommand{\ds}{\displaystyle}

\newcommand{\mrd}{\mathrm{d}}
\newcommand{\mre}{\mathrm{e}}
\newcommand{\mri}{\mathrm{i}}

\newcommand{\bd}{{\bf d}}

\newcommand{\bx}{{\bf x}}
\newcommand{\bu}{{\bf u}}

\newcommand{\bphi}{\mbox{\boldmath$\phi$}}

\newcommand{\bpsi}{\mbox{\boldmath$\psi$}}

\newcommand{\DefinedAs}[0]{\mathrel{\mathop:}=}

\DeclareMathOperator*{\logdet}{log\,det}

\DeclareMathOperator*{\minimize}{minimize}

\DeclareMathOperator*{\subject}{subject~to}
\DeclareMathOperator*{\rank}{rank}

\definecolor{bgblue}{rgb}{0.04,0.19,0.53}
\definecolor{dblue1}{rgb}{0,0.3,0.7}
\definecolor{dred}{rgb}{0.4,0.2,0}

\newtheorem{remark}{Remark}

\newcommand{\enma}[1]   {\ensuremath{#1}}

\newcommand{\beq}{\begin{equation}}
\newcommand{\eeq}{\end{equation}}
\newcommand{\bseq}{\begin{subequations}}
\newcommand{\eseq}{\end{subequations}}
\newcommand{\beqn}{\begin{eqnarray}}
\newcommand{\eeqn}{\end{eqnarray}}
\newcommand{\ba}{\begin{array}}
\newcommand{\ea}{\end{array}}
\newcommand{\bct}{\begin{center}}
\newcommand{\ect}{\end{center}}
\newcommand{\btmz}{\begin{itemize}}
\newcommand{\etmz}{\end{itemize}}
\newcommand{\benum}{\begin{enumerate}}
\newcommand{\eenum}{\end{enumerate}}

\newcommand{\norm}[1]{\| #1 \|}                 %does not make large \|

\newcommand{\diag}      {\enma{\mathrm{diag}}}

\newcommand{\trace}     {\enma{\mathrm{trace}}}

\newcommand{\bv}{{\bf v}}

\newcommand{\matbegin}{
        \left[
}
\newcommand{\matend}{
        \right]
}

\newcommand{\tbo}[2]{
  \matbegin \begin{array}{c}
       #1 \\ #2
       \end{array} \matend }

\newcommand{\obt}[2]{
  \matbegin \begin{array}{cc}
       #1 & #2
       \end{array} \matend }

\newcommand{\tbt}[4]{
  \matbegin \begin{array}{cc}
       #1 & #2 \\ #3 & #4
       \end{array} \matend }

\newcommand{\be}{\begin{equation}}
\newcommand{\ee}{\end{equation}}

\newcommand{\cplxs}{ C\kern -.35em \rule{0.03 em}{.7 ex}~   }

\def\complex{\hbox{C\kern -.45em \rule{0.03 em}{1.5 ex}}~}

\newcommand{\bi}{\begin{itemize}}
\newcommand{\ei}{\end{itemize}}

\newcommand{\bk}{{\bf{k}}}
\newcommand{\bw}{{\bf{w}}}
\newcommand{\bE}{{\bf E}}

\jname{Annu. Rev. Control Robot. Auton. Syst.}
\jvol{3}
\jyear{2019}
\doi{10.1146/((please add article doi))}

\begin{document}

% Page header
\markboth{Zare, Georgiou, Jovanovi\'c}{Stochastic dynamical modeling of turbulent flows}

% Title
\title{Stochastic dynamical modeling of turbulent flows}

%Authors, affiliations address.
\author{\mbox{A.\ Zare$^1$, T.\ T.\ Georgiou$^2$, and M.\ R.\ Jovanovi\'c$^3$}
\affil{$^1$Department of Mechanical Engineering, University of Texas at Dallas, Richardson, Texas 75080, USA}
\affil{$^2$Department of Mechanical and Aerospace Engineering, University of California, Irvine, California 92697, USA}
\affil{$^3$Ming Hsieh Department of Electrical and Computer Engineering, University of Southern California, Los Angeles, California 90089, USA; email: mihailo@usc.edu}
}

%Abstract
\begin{abstract}
Advanced measurement techniques and high performance computing have made large data sets available for a wide range of turbulent flows that arise in engineering applications. Drawing on this abundance of data, dynamical models can be constructed to reproduce structural and statistical features of turbulent flows, opening the way to the design of effective model-based flow control strategies. This review describes a framework for completing second-order statistics of turbulent flows by models that are based on the Navier-Stokes equations linearized around the turbulent mean velocity. Systems theory and convex optimization are combined to address the inherent uncertainty in the dynamics and the statistics of the flow by seeking a suitable parsimonious correction to the prior linearized model. Specifically, dynamical couplings between states of the linearized model dictate structural constraints on the statistics of flow fluctuations. Thence, colored-in-time stochastic forcing that drives the linearized model is sought to account for and reconcile dynamics with available data (i.e., partially known second order statistics). The number of dynamical degrees of freedom that are directly affected by stochastic excitation is minimized as a measure of model parsimony. The spectral content of the resulting colored-in-time stochastic contribution can alternatively be seen to arise from a low-rank structural perturbation of the linearized dynamical generator, pointing to suitable dynamical corrections that may account for the absence of the nonlinear interactions in the linearized model.
\end{abstract}

%Keywords, etc.
\maketitle

% Introduction
\section{INTRODUCTION}

Turbulent flows are at the center of many key processes in nature and in engineering applications. Energy dissipation caused by turbulent fluctuations around airplanes, ships, and submarines increases resistance to motion (i.e., skin-friction drag) and fuel consumption and compromises the performance of vehicles. This motivates the design of flow control strategies for the improved performance of air and water vehicles and other systems that involve turbulent flows~\cite{jos98,gad00}. 

Models that are based on the Navier-Stokes (NS) equations capture the dynamics and statistical features of fluid flows. However, these models are given by 3D nonlinear PDEs and they involve a number of degrees of freedom that is prohibitively large for analysis and control synthesis~\cite{chomoi12,slokhoalodargrolurmav13}. Moreover, to this day, a detailed understanding of the mechanisms responsible for the dissipation of energy in turbulent flows is missing. As a result, traditional flow control techniques are largely empirical, and they rely on physical intuition, numerical simulations, and experiments. Even though these provide invaluable insights, they are costly, time-consuming, and are not suitable for model-based controller design.

Direct numerical simulation (DNS) offers a computational approach to finding a solution to the NS equations. At moderate Reynolds numbers DNS provides important insight into structural and statistical features of turbulent flows but computational complexity increases roughly as the cube of the Reynolds number and DNS becomes prohibitively expensive in most flow regimes that are encountered in engineering practice~\cite{chomoi12}. An alternative to DNS has been to either fully resolve large-scale 3D turbulent flow structures and to model the impact of smaller scales or to focus on statistical signatures of turbulent flows, i.e., the mean flow components and their higher-order moments. The former approach gives rise to large-eddy simulation (LES) which relies on modeling the impact of small unresolved physical scales~\cite{sag06}, and the latter forms the basis for statistical theory of turbulence~\cite{wil98}. While LES accurately captures large-scale unsteady motions that dominate flows around air and water vehicles, its computational cost is still too high for it to be incorporated into aerodynamic design~\cite{slokhoalodargrolurmav13}. Since an exact set of dynamical equations that govern the evolution of statistics of turbulent flows does not exist, the statistical theory of turbulence aims to develop approximate mathematical models for turbulent flows~\cite{durrei11}. Indeed, recent research suggests that conventional techniques can be significantly enhanced using low-complexity models that are convenient for real-time control design and optimization~\cite{kimbew07}. 

In general, modeling can be seen as an inverse problem where a search in parameter space aims to identify a parsimonious representation of data. For turbulent flows, the advent of advanced measurement techniques and high performance parallel computing has resulted in large data sets for a wide range of flow configurations and speeds. Tapping on this abundance of data, dynamical models can be constructed to reproduce structural and statistical features of turbulent flows. 

The prevalence of coherent structures in turbulent wall-bounded shear flows~\cite{rob91,adr07,smimckmar11,jim18} has inspired the development of data-driven techniques for reduced-order modeling of turbulent flows~\cite{row05,lum07,sch10,jovschnicPOF14,rowdaw17,towschcol18}. However, unreliable measurements and data anomalies challenge a sole reliance on data as such models are agnostic to the underlying physics. Furthermore, control actuation and sensing may significantly alter the identified modes in unpredictable ways. This compromises the performance of data-driven models in regimes that were not accounted for in the training process and introduces nontrivial challenges for model-based control design~\cite{noamortad11,tadnoa11}. A promising alternative is to leverage the underlying physics in the form of a prior model that arises from first principles, e.g., linearization of the NS equations around stable flow states. The subject of this review is to highlight recent developments in combining data-driven techniques with systems theory and optimization to enhance predictive capabilities of physics-based dynamical models.

Over the last three decades, important dynamical aspects of transitional and turbulent flows have been captured by the analysis of the linearized NS equations. Specifically, the non-normality of the linearized dynamical generator introduces interactions among exponentially decaying normal modes~\cite{tretrereddri93,sch07}. This property has been used to explain high flow sensitivity in the early stages of transition and to identify key mechanisms for subcritical transition to turbulence; even in the absence of modal instability, bypass routes to transition can be triggered by large transient growth~\cite{gus91,butfar92,redhen93,henred94,schhen94} or large amplification of deterministic and stochastic disturbances~\cite{tretrereddri93,farioa93,bamdah01,mj-phd04,jovbamJFM05,ranzarhacjovPRF19b}. Similar amplification mechanisms have been observed for the linearized NS equations around the turbulent mean velocity~\cite{mj-phd04,butfar93,farioa93a,farioa98,mcksha10}. Additional insights into the geometric scaling of dominant modes over various flow conditions have been provided by low-order representations resulting from singular value decomposition of the associated frequency response operator~\cite{mcksha10,moashatromck13,moajovtroshamckPOF14}.

The nonlinear terms in the NS equations play an important role in the growth of flow fluctuations, transition to turbulence, and in sustaining turbulent flow. Since these terms are conservative, they do not contribute to the transfer of energy between the mean flow and velocity fluctuations, but they transfer energy between different spatio-temporal Fourier modes~\cite{mcc91,durrei11}. This feature has inspired modeling the effect of nonlinearity using additive forcing to the linearized equations that govern the dynamics of fluctuations. Early efforts in this direction focused on modeling homogeneous isotropic turbulence~\cite{kra59,kra71,ors70,monyag75}. Stochastically-forced linearized NS equations were later used to model heat and momentum fluxes as well as spatio-temporal spectra in quasi-geostrophic turbulence~\cite{farioa93c,farioa94a,delfar95}, while structural features of wall-bounded turbulent flows were captured using the spatio-temporal frequency responses of the linearized NS equations~\cite{farioa93,bamdah01,mj-phd04,jovbamJFM05,mcksha10,hwacosJFM10a,hwacosJFM10b,jovbamCDC01}. In these studies, forcing is used to model exogenous excitation sources and uncertain initial conditions, as well as to replicate the effects of the nonlinear terms in the full NS equations.

This review explains how stochastic dynamical models can enhance the linearized NS equations so as to accurately replicate observed statistical features of turbulent flows. This is accomplished by bringing together tools from systems theory and convex optimization in order to suitably shape the power spectrum of additive stochastic forcing into the dynamical equations. We focus on replicating second-order statistics and cast the corresponding model identification as a convex optimization problem. The resulting stochastic component can be linked to a structural (low-rank) perturbation of the dynamical generator suggesting suitable correction to account for the absence of the nonlinear interactions.

The review is organized as follows. In Section~\ref{sec.NS-stats}, we provide the background on the NS equations and turbulence modeling. In Section~\ref{sec.StochasticallyForcedLNS}, we introduce the stochastically-forced linearized NS equations and describe structural constraints on admissible state covariances and input power spectra. In Section~\ref{sec.CCP}, we demonstrate the necessity for colored-in-time stochastic forcing and formulate a convex optimization problem aimed at matching available and completing unavailable second-order statistics of turbulent flows via low-complexity stochastic dynamical models. In Section~\ref{sec.channel}, we apply the stochastic modeling approach of Section~\ref{sec.CCP} to a turbulent channel flow, verify its utility in linear stochastic simulations, and examine the resulting spatio-temporal spectrum. We close the paper by discussing the outstanding research issues and provide concluding remarks in Section~\ref{sec.conclusions}.

	\vspace*{-2ex}
\section{THE NAVIER-STOKES EQUATIONS}
\label{sec.NS-stats}

Flows of incompressible Newtonian fluids are governed by the NS and continuity equations,
\begin{align}
	\label{eq.NScts}
	\ba{rcl}
    	\partial_t \bu
	\; + \; 
	(\bu \cdot \nabla) \bu
    	&\; = \;&
    	-
    	\nabla P
    	\; + \; 
	\dfrac{1}{Re} \, \Delta \bu,
    	\\
    	0
    	&\; = \;&
    	\nabla \cdot \bu,
	\ea
\end{align}
where $\bu$ is the velocity vector which satisfies the no-slip and no-penetration boundary conditions at a stationary solid surface, $P$ is the pressure, $\nabla$ and $\Delta = \nabla \cdot \nabla$ are the gradient and Laplacian operators, and $\partial_t$ is the partial derivative with respect to time. The NS equations are nonlinear PDEs in spatial coordinates $\bx$ and time $t$ and the continuity equation reflects the static-in-time divergence-free requirement on the velocity field. The flow is parameterized by the Reynolds number which determines the ratio of inertial to viscous forces, $Re \DefinedAs \bar{u} h/\nu$, where $\bar{u}$ and $h$ are the characteristic velocity and length of the flow and $\nu$ is the kinematic viscosity. Spatial coordinates in Equation~\ref{eq.NScts} are non-dimensionalized by $h$, velocity by $\bar{u}$, time by $h/\bar{u}$, and pressure by $\rho \bar{u}^2$, where $\rho$ is the~fluid~density. 

	\vspace*{-2ex}
\subsection*{Mean flow equations and the closure problem}
		
When the flow becomes turbulent, it reaches a statistically stationary state in which variables still vary in time but their statistics are time-independent. To analyze the statistical properties of the flow, the velocity and pressure fields are decomposed into the sum of the turbulent mean components ($\bar{\bu},\bar{P}$) and fluctuations ($\bv,p$) around them,
\be
	\ba{rcl}
	(\bu,P) 
	& = &
	( \bar{\bu} + \bv, \bar{P} + p ),
	~~
	(\bar{\bu}, \bar{P} )
	\; = \;
	( \left< \bu \right>, \left< P \right> ),
	~~
	( \left< \bv \right>,  \left< p \right> )
	\; = \;
	(0, 0),
	\ea
\non
\ee
where $\left<\cdot\right>$ denotes the time-average operator, e.g., 
\begin{align*}
	\left<\bu(\bx,t) \right>
	\;=\;
	\lim_{T \, \to \, \infty} \dfrac{1}{T} \int^T_0 \bu(\bx,t\,+\,\tau)\, \mrd \tau.
\end{align*}
Averaging Equation~\ref{eq.NScts} yields the Reynolds-averaged NS equations~\cite{mcc91,durpet00,pop00},
\begin{align}
	\label{eq.RANS}
	\ba{rcl}
	\partial_t \bar{\bu}
	\; + \; 
		\left(\bar{\bu} \cdot \nabla \right) \bar{\bu}
		& \; = \; &
		- 
		\nabla \bar{P} 
       		\; + \;
       		\dfrac{1}{Re} \, \Delta \bar{\bu}
        		\; - \;
		\nabla \cdot \left<\bv \bv^T \right>,
       		\\
       		0 
    		& \; = \; &
    		\nabla \cdot \bar{\bu},
	\ea
\end{align}
which govern the evolution of the turbulent mean profiles ($\bar{\bu},\bar{P}$). Relative to the NS equations~\ref{eq.NScts}, the mean-momentum equations~\ref{eq.RANS} contain one additional term which depends on the second order moment of the velocity fluctuation vector $\bv$, $\left<\bv \bv^T \right>$. This symmetric tensor arises from momentum transfer by the velocity fluctuations and it has profound influence on the mean flow quantities and thereby on our ability to \mbox{predict the skin-friction drag~\cite{mcc91}.} 

For a three-dimensional flow, Equation~\ref{eq.RANS} consists of four independent equations governing the dynamics of the mean velocity and pressure fields ($\bar{\bu},\bar{P}$). However, these equations contain more than four unknowns; in addition to $\bar{\bu}$ and $\bar{P}$, the {\em Reynolds stresses\/} $\left<\bv \bv^T \right>$ are also unknown. This is a consequence of a {\em closure problem\/} that cannot be resolved in the absence of additional information about the second-order statistics of velocity fluctuations. Since the NS equations are nonlinear, the $n$th velocity moment depends on the $(n+1)$th moment~\cite{mcc91} making it challenging to determine such second-order statistics. 

Statistical theories of turbulence attempt to overcome the closure problem by modeling the effect of the Reynolds stress tensor on the mean flow quantities rather than explicitly resolving the nonlinear terms~\cite{mcc91,durpet00,pop00}. This is typically achieved by expressing higher-order moments in terms of the lower-order moments via a combination of physical intuition and empirical observations with rigorous approximation of the flow equations; see {\bf Figure~\ref{fig.equilibrium1}}. For example, the turbulent viscosity hypothesis seeks approximate solutions of Equation~\ref{eq.RANS} by relating turbulent stresses to mean velocity gradients via~\cite{pop00},
\begin{align*}
	\left< \bv \bv^T \right> 
	\;-\; 
	\dfrac{1}{3}\, \trace \left(\left< \bv \bv^T \right>\right) I
	\;=\;
	-\dfrac{\nu_{T}}{Re} \left( \nabla \bar{\bu} 
	\;+\; 
	\left(\nabla \bar{\bu} \right)^T\right),
\end{align*}
where $\nu_T$ is the turbulent viscosity and $I$ is the identity tensor. Unfortunately, a general purpose expression for $\nu_T$ does not exist and turbulence models are required to relate it to other flow quantities, e.g., second-order statistics of the velocity fluctuations.

With appropriate choices of velocity and length scales, turbulent viscosity can be expressed as~\cite{pop00}
\begin{align*}
	\nu_T \;=\; c Re^2 ({k^2}/{\eps}),
\end{align*}
where $k$ and $\eps$ denote the turbulent kinetic energy and its rate of dissipation and $c$ is the constant. The $k$-$\eps$ model~\cite{jonlau72,lausha74} provides two differential transport equations for computing $k$ and $\eps$ and it is widely used in commercial computational fluid dynamics codes and in engineering practice. Even though these are less complex than the NS equations, they are still computationally expensive, produce reliable result only for certain flow configurations, and are not convenient for control design and optimization; see~\cite{pop00} for additional details. In what follows, we describe an alternative approach to turbulence modeling, which approximates the Reynolds stresses using the second-order statistics of the stochastically-forced NS equations linearized around the turbulent mean flow; see {\bf Figure~\ref{fig.equilibrium2}}. We also demonstrate how second-order statistics resulting from DNS and experiments can be used to refine the predictive capability of  models that arise from first principles.

\begin{figure}
\begin{center}
\begin{tabular}{cc}
\hspace{-.4cm}
\subfigure[]{\label{fig.equilibrium1}
         %_______________________________________________________________________________
%
%   tikz figure for inclusion in exercises:
%
%   clp_2dof_output_pert_config:  closed loop system with perturbed models.  This uses
%                                              and input perturbation and the 2-DOF structure used for
%                                              the satellite example.
%
%   Marcello Colombino, 4 May 2015
%  
%_______________________________________________________________________________
%
% TikZ styles for drawing
%
\input{figures/Tikz_common_styles}
%
%   set a filename for externalization
% \tikzsetnextfilename{clp_2dof_input_pert_config}
%
\noindent
\begin{tikzpicture}[scale=1, auto, >=stealth']
  
    \footnotesize

   % specify with respect to the plant
    
     \node[block, minimum height = 1.2cm, top color=Red!10, bottom color=Red!10] (sys1) {$\ba{c} \mbox{mean flow} \\ \mbox{equations}\ea$};
     
     \node[block, minimum height = 1.2cm, top color=RoyalBlue!10, bottom color=RoyalBlue!10] (sys2) at ($(sys1.east) + (3.1cm,0cm)$) {$\ba{c} \mbox{turbulence} \\ \mbox{model}\ea$};
          
%     \node[] (input-node) at ($(sys2.north) + (0,1.2cm)$) {$\ba{c} \mbox{stochastic} \\ \mbox{forcing}\ea$}; 
     
%     \node[] (output-node) at ($(sys1.north) + (0,1.2cm)$) {$\ba{c} \mbox{turbulent} \\ \mbox{drag}\ea$};
     
     \node[] (mid-node1) at ($(sys2.center) + (1.25cm,0cm)$) {};
     
     \node[] (mid-node2) at ($(sys2.center) + (1.25cm,-1.5cm)$) {};
     
     \node[] (mid-node3) at ($(sys1.center) - (0cm,1.5cm)$) {};
     
    % now link the nodes
		
    \draw [connector] (sys1.east) -- node [midway, above] {$\ba{c} \mbox{turbulent} \\ \mbox{mean velocity}\ea$} (sys2.west);
    
    \draw [line] (sys2.east) -- (mid-node1.center);
    
    \draw [line] (mid-node1.center) -- (mid-node2.center);
    
     \draw [line] (mid-node2.center) -- node [midway, above] {$\ba{c} \mbox{second-order} \\ \mbox{statistics}\ea$} (mid-node3.center);

    \draw [connector] (mid-node3.center) -- (sys1.south);
    
%    \draw [connector] (sys1.north) -- (output-node);
    
%    \draw [connector] (input-node) -- (sys2.north);
    
%    \draw [connector] (mid-node1) -- node [midway, right] {$f_\eta$} (esum2.north);
%    	
%    \draw [line] (sys3.east) -- (R);
%    \draw [connector] (R.west) -- node [midway, above] {$\eta$} (output-node);
%%    \draw [connector] (esum1.east) -- node [midway, below] {$\bv$} (output-node);
%    
%    \draw [connector] (R.south) |- (sys4.east);
%    
%    \draw [connector] (sys4.west) -| (esum1.south);
%    
%    \draw [connector] (esum1.east) -- (sys1.west);
                                          
\end{tikzpicture}
%_______________________________________________________________________________
         }
         &
         \hspace{-.7cm}
         \subfigure[]{\label{fig.equilibrium2}
         %_______________________________________________________________________________
%
%   tikz figure for inclusion in exercises:
%
%   clp_2dof_output_pert_config:  closed loop system with perturbed models.  This uses
%                                              and input perturbation and the 2-DOF structure used for
%                                              the satellite example.
%
%   Marcello Colombino, 4 May 2015
%  
%_______________________________________________________________________________
%
% TikZ styles for drawing
%
\input{figures/Tikz_common_styles}
%
%   set a filename for externalization
% \tikzsetnextfilename{clp_2dof_input_pert_config}
%
\noindent
\begin{tikzpicture}[scale=1, auto, >=stealth']
  
    \footnotesize

   % specify with respect to the plant
    
     \node[block, minimum height = 1.2cm, top color=Red!10, bottom color=Red!10] (sys1) {$\ba{c} \mbox{mean flow} \\ \mbox{equations}\ea$};
     
     \node[block, minimum height = 1.2cm, top color=RoyalBlue!10, bottom color=RoyalBlue!10] (sys2) at ($(sys1.east) + (3.1cm,0cm)$) {$\ba{c} \mbox{linearized} \\ \mbox{dynamics}\ea$};
          
     \node[] (input-node) at ($(sys2.north) + (0,1cm)$) {$\ba{c} \mbox{stochastic} \\ \mbox{forcing}\ea$}; 
     
%     \node[] (output-node) at ($(sys1.north) + (0,1.2cm)$) {$\ba{c} \mbox{turbulent} \\ \mbox{drag}\ea$};
     
     \node[] (mid-node1) at ($(sys2.center) + (1.25cm,0cm)$) {};
     
     \node[] (mid-node2) at ($(sys2.center) + (1.25cm,-1.5cm)$) {};
     
     \node[] (mid-node3) at ($(sys1.center) - (0cm,1.5cm)$) {};
     
    % now link the nodes
		
    \draw [connector] (sys1.east) -- node [midway, above] {$\ba{c} \mbox{turbulent} \\ \mbox{mean velocity}\ea$} (sys2.west);
    
    \draw [line] (sys2.east) -- (mid-node1.center);
    
    \draw [line] (mid-node1.center) -- (mid-node2.center);
    
     \draw [line] (mid-node2.center) -- node [midway, above] {$\ba{c} \mbox{second-order} \\ \mbox{statistics}\ea$} (mid-node3.center);

    \draw [connector] (mid-node3.center) -- (sys1.south);
    
%    \draw [connector] (sys1.north) -- (output-node);
    
    \draw [connector] (input-node) -- (sys2.north);
    
%    \draw [connector] (mid-node1) -- node [midway, right] {$f_\eta$} (esum2.north);
%    	
%    \draw [line] (sys3.east) -- (R);
%    \draw [connector] (R.west) -- node [midway, above] {$\eta$} (output-node);
%%    \draw [connector] (esum1.east) -- node [midway, below] {$\bv$} (output-node);
%    
%    \draw [connector] (R.south) |- (sys4.east);
%    
%    \draw [connector] (sys4.west) -| (esum1.south);
%    
%    \draw [connector] (esum1.east) -- (sys1.west);
                                          
\end{tikzpicture}
%_______________________________________________________________________________
         }
\end{tabular}
\end{center}
\caption{(a) Conventional turbulence models are used to compute second-order statistics which drive the mean flow equations. (b) An alternative approach utilizes stochastically-forced linearized dynamics around the turbulent mean velocity to compute the second-order statistics of velocity fluctuations.}
\label{fig.equilibrium}
\end{figure}

\begin{figure}
\begin{center}
\begin{tabular}{rcrc}
        \subfigure[]{\label{fig.channel}}
        &&
        \hspace{-3cm}
        \subfigure[]{\label{fig.meanflow}}
        &
        \\[-.4cm]
        &
        \hspace{-4cm}
        \begin{tabular}{c}
            	\vspace{.2cm}
         	\includegraphics[height=2.8cm]{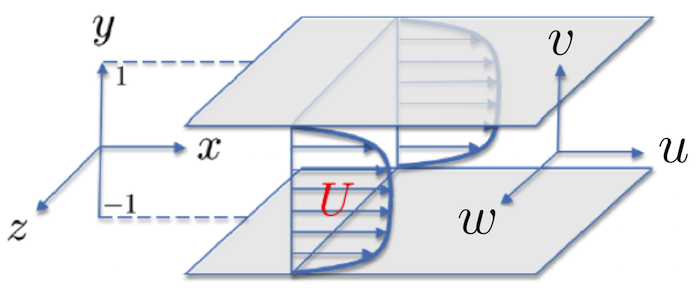}
	\end{tabular}
	\hspace{-3.2cm}
         &&
         \hspace{-1cm}
         \begin{tabular}{rl}
            	\begin{tabular}{c}
            		\vspace{.6cm}
            		\normalsize{\rotatebox{90}{$y$}}
            	\end{tabular}
            	&
		\hspace{-4.9cm}
            	\begin{tabular}{c}	
            	         \includegraphics[height=4cm]{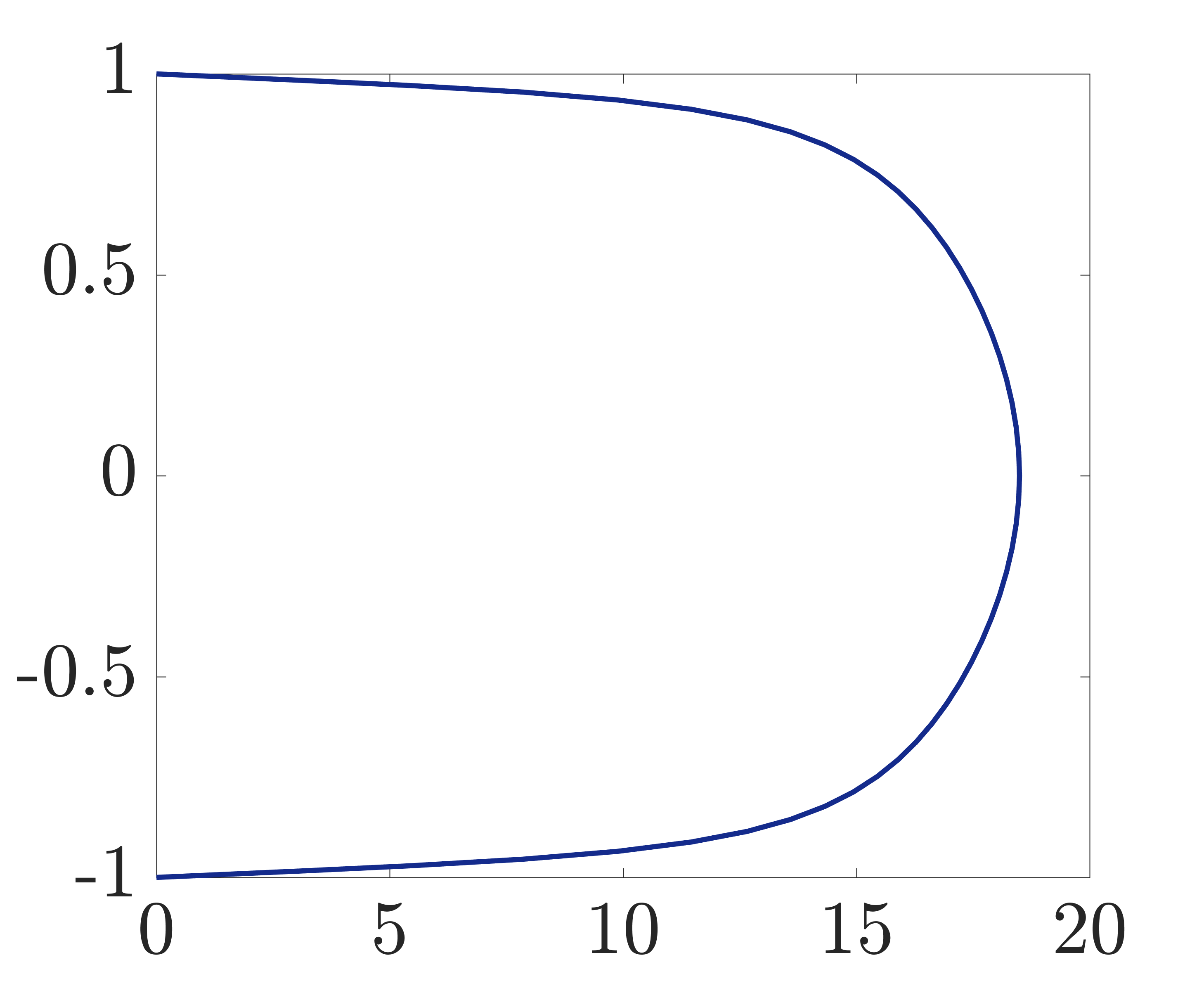}
            	         \\
            	         $U(y)$
            	\end{tabular}
	\end{tabular}
\end{tabular}
\end{center}
\caption{(a) Geometry of a pressure-driven turbulent channel flow between two parallel infinite walls. (b) Turbulent mean velocity profile $U(y)$ in a flow with friction Reynolds number $Re = 186$.}
\label{fig.channel-meanflow}
\end{figure}

	\vspace*{-2ex}
\section{\mbox{STOCHASTICALLY-FORCED LINEARIZED NAVIER-STOKES EQUATIONS}}
\label{sec.StochasticallyForcedLNS}

The dynamics of small velocity and pressure fluctuations ($\bv,p$) around the turbulent mean profile ($\bar{\bu},\bar{P}$) are governed by the linearized NS and continuity equations 
\begin{subequations}
	\label{eq.turblin}
	\begin{eqnarray}
    		\partial_t \bv
		\; + \;
		\left( \nabla \cdot \bar{\bu} \right) \bv
     		\; + \; 
    		\left( \nabla \cdot \bv \right) \bar{\bu}
    		& \; = \; &
    		-
    		\nabla p
    		\; + \;
    		\dfrac{1}{Re} \, \Delta \bv
    		\; + \;
    		\bd,
    		\\[-0.1cm]
		\label{eq.continuity}
    		0
    		& \; = \; &
    		\nabla \cdot \bv.
	\end{eqnarray}
\end{subequations}
Here, $\bd$ represents an additive zero-mean stationary stochastic input that triggers a statistical response of the linearized dynamics. In what follows, we describe how available second-order statistics of turbulent channel flows can be reproduced using the stochastically-forced model represented by Equation~\ref{eq.turblin} and a suitable choice of power spectrum for the input $\bd$. While we focus on the turbulent channel flow, it is noteworthy that the reviewed methodology and theoretical framework are applicable to other flow configurations.

In strongly inertial regimes, all flows transition to turbulence and a channel flow, with geometry shown in {\bf Figure~\ref{fig.channel}}, is commonly used as a benchmark for modeling, analysis, and control of wall-bounded turbulence. As illustrated in {\bf Figure~\ref{fig.meanflow}}, the turbulent mean velocity in channel flow only contains a streamwise component $\bar{\bu} = [\,U(y)\,~0\,~0\,]^T$, and the linearized model that governs the dynamics of velocity fluctuations $\bv \DefinedAs [\,u\,\;v\,\;w\,]^T$, in the streamwise, $x$, wall-normal, $y$, and spanwise, $z$, directions takes the~form, 
	 \begin{equation}
    \ba{rcccrcl}
    \partial_{t} u & + & U (y) \, \partial_{x} u & + & U' (y) \, v
    & \; = \; &
    - \, \partial_{x} p \; + \;  (1/Re) \, \Delta u  \; + \;  d_u
    \\[.15cm]
    \partial_{t} {v} & + & U (y) \, \partial_{x} v  & &
    & \; = \; &
    - \, \partial_{y} p  \; + \;  (1/Re) \, \Delta v \; + \;  d_v
    \\[.15cm]
    \partial_{t} {w} & + & U (y) \, \partial_{x} w  & &
    & \; = \; &
    - \, \partial_{z} p \; + \; (1/Re) \, \Delta w \; + \; d_w
    \\[.15cm]
    & & & & 0
    & \; = \; &
    \phantom{-} \, \partial_x u \; + \; \partial_y v \; + \; \partial_z w.
    \ea
    \label{eq.linNSforced}
    \end{equation}
Here, $U' (y) \DefinedAs \mrd U (y)/ \mrd y$ and $\bd \DefinedAs [\,d_u\,\;d_v\,\;d_w\,]^T$ is the body forcing fluctuation vector. By selecting the channel half-height $h$ and the friction velocity $u_\tau$ as the proper scales, the flow is characterized by the friction Reynolds number $Re \DefinedAs u_\tau h/\nu$. 

The linearized dynamics in Equation~\ref{eq.linNSforced} are time-invariant and have constant coefficients in the wall-parallel directions; thus, the Fourier transform in $x$ and $z$ can be used to obtain a one-dimensional system of PDEs (in $y$ and $t$) parameterized by the horizontal wavenumbers $\bk \DefinedAs (k_x,k_z)$. Furthermore, a standard conversion can be used to eliminate the pressure from the equations and bring the descriptor form in Equation~\ref{eq.linNSforced} into the form of an evolution model in which the state is determined by the wall-normal velocity, $v$, and vorticity, $\eta = \partial_z u - \partial_x w$, fluctuations~\cite{kimmoimos87,jovbamJFM05} with the boundary conditions $v(y = \pm1, \bk, t) = \partial_y v(y = \pm1, \bk, t) = \eta(y = \pm1, \bk, t) = 0$. A pseudo-spectral technique~\cite{weired00} with $N$ collocation points in $y$ approximates the underlying operators and a change of variables described in~\cite[Appendix A]{zarjovgeoJFM17} is used to obtain a finite-dimensional state-space representation in which the energy of velocity fluctuations at any $\bk$ is determined by the Euclidean norm of the state vector $\bpsi \DefinedAs [\,v\,\;\eta\,]^T$,
\begin{align}
	\label{eq.lnse1}
        \ba{rcl}
            \dot{\bpsi} (\bk, t)
            & \; = \; &
            A (\bk)
            \,
            \bpsi (\bk, t)
            \; + \;
            B (\bk)
            \,
            \bd (\bk, t),
            \\[0.1cm]
            \bv (\bk, t)
            & \; = \; &
            C (\bk)
            \,
            \bpsi (\bk, t).
        \ea
\end{align}
Here, $\bpsi(\bk, t)  \in \bbC^{2N}$, $\bd (\bk, t) \in \bbC^{3N}$ is the input vector, $\bv (\bk, t) \in \bbC^{3N}$ is the velocity fluctuation vector, the matrix $A(\bk)$ determines dynamical interactions between the state variables, $B (\bk)$ specifies the way the input $\bd (\bk, t)$ enters into the evolution model, and the output matrix $C (\bk)$ relates the state vector $\bpsi (\bk, t)$ to the velocity fluctuation vector $\bv (\bk, t)$. 
 
%\subsection{Sidebars and Margin Notes}
%% Margin Note
\begin{marginnote}[]
The evolution model is obtained from Equation~\ref{eq.linNSforced} as follows. Applying the divergence operator $\nabla$ to the linearized NS equations yields an expression for $\Delta p$. The equation for $v$ is obtained by acting with the Laplacian $\Delta$ on the second equation in Model~\ref{eq.linNSforced} and utilizing the expression for $\Delta p$ to eliminate the pressure $p$. The equation for $\eta$ is obtained by taking the curl of the linearized NS equations. This yields two PDEs that govern the evolution of $v$ and $\eta$ that involve only $v$, $\eta$, and $\bd$.
\end{marginnote}

\begin{marginnote}[]
The linearized NS equations around the turbulent mean velocity profile are stable~\cite{mal56,reytie67}, i.e., all eigenvalues of $A$ in Equation~\ref{eq.lnse1} are in the left-half of the complex plane. 
\end{marginnote}

	\vspace*{-2ex}
\subsection{Algebraic relations between input and state statistics}
\label{sec.alg-relations}

In channel flow, the NS equations linearized around the turbulent mean flow are stable~\cite{mal56,reytie67} and the steady-state covariance matrix $X (\bk)$ of the state vector in Equation~\ref{eq.lnse1} 
	\begin{align}
	X (\bk)
	\;=\;
	\lim_{t \, \to \, \infty}
	\bE( 
	\bpsi(\bk,t)\, \bpsi^*(\bk,t) 
	),
	\label{eq.statecovariance}
\end{align}
satisfies the Lyapunov-like equation~\cite{geo02b,geo02a}
\begin{align}
	A (\bk) \, X (\bk)
	\,+\, 
	X (\bk) \,A^* (\bk) 
	\;=\; 
	-\,B (\bk) \,H^* (\bk)
	\,-\, 
	H (\bk) \,B^* (\bk),
	\label{eq.lyap_BH}
\end{align}
where $\bE$ is the expectation operator and $*$ is complex-conjugate-transpose. For colored-in-time $\bd (\bk,t)$, $H (\bk)$ is a matrix that quantifies the cross-correlation between the input and the state in Equation~\ref{eq.lnse1} in statistical steady-state~\cite[Appendix B]{zarjovgeoJFM17}, 
\begin{align*}
	H(\bk) 
	\;=\;
	\lim_{t \, \to \, \infty} \bE\,( \bpsi(\bk,t) \bd^*(\bk,t)) 
	\;+\;
	\dfrac{1}{2}\, B(\bk)\, \Omega(\bk).
\end{align*}
When the input $\bd (\bk,t)$ in Equation~\ref{eq.lnse1} is zero-mean and white-in-time with covariance matrix $\Omega (\bk)$, i.e., $\bE ( \bd(\bk,t) ) = 0$ and $\bE ( \bd(\bk,t)\, \bd^*(\bk,\tau) ) = \Omega(\bk) \delta (t - \tau)$, $H (\bk)$ simplifies to $H(\bk)=(1/2) B(\bk) \Omega(\bk)$ and Equation~\ref{eq.lyap_BH} reduces to the standard algebraic Lyapunov equation, 
\begin{align}
	A(\bk)\, X (\bk)
	\;+\;
	X(\bk)\, A^*(\bk) 
	\;=\; 
	-B(\bk) \, \Omega(\bk) B^*(\bk).
	\label{eq.standard_lyap}
\end{align} 
The steady-state velocity covariance matrix $V (\bk)$ can be obtained from $X (\bk)$, 
\begin{align}
	V (\bk)
	\;=\;
	\lim_{t \, \to \, \infty}
	\bE( 
	\bv(\bk,t)\, \bv^*(\bk,t) 
	)
	\; = \;
	C (\bk)\, X (\bk)\, C^* (\bk).
	\label{eq.Xphi_relation}
\end{align}
Since the dynamics are parameterized by wavenumbers $\bk$, the entries of $V(\bk)$ determine two-point correlations of velocity fluctuations in the wall-normal direction $y$; see~\cite{moimos89}.

		\vspace*{-2ex}
\subsection{Spatio-temporal correlations}
\label{sec.spatio-temporal-corr}

At any $\bk$, the matrix $V (\bk)$ determines two-point correlations in the wall-normal direction of velocity fluctuations in statistical steady-state and the lagged covariance matrix,
\begin{align}
\label{eq.correlation-tensor}
	R_{\bv \bv}(\bk, \tau)
	\;\DefinedAs\;
	\lim_{t \, \to \, \infty} 
	\bE\, ( \bv(\bk, t)\,\bv^*(\bk, t + \tau) ),
\end{align}
captures spatio-temporal correlations. Furthemore, the application of the temporal Fourier transform yields the spectral density matrix $S_{\bv \bv}(\bk,\omega)$ of the output $\bv(\bk,t)$,
\begin{align}
\label{eq.CSD}
	\ba{rcl}
	S_{\bv \bv}(\bk, \omega)
	& = &
	\ds{\int^{+\infty}_{-\infty}} R_{\bv \bv}(\bk, \tau) \, \mre^{-\mri \omega \tau} \mrd \tau,
	\ea
\end{align}
which parameterizes two-point velocity correlations across wavenumbers $\bk$ and temporal frequencies $\omega$. The matrix $S_{\bv \bv}(\bk,\omega)$ can be expressed in terms of the spectral density matrix $S_{\bd \bd}(\bk,\omega)$ of the input $\bd(\bk,t)$, 
	\begin{align*}
	S_{\bv \bv}(\bk,\omega)
	\; = \;
	T_{\bv \bd} (\bk, \omega)\, S_{\bd \bd}(\bk,\omega) \,T^*_{\bv \bd} (\bk,\omega),
\end{align*}
where $T_{\bv \bd}(\bk,\omega)$ is the spatio-temporal frequency response of the LTI system in Equation~\ref{eq.lnse1},
\begin{align}
	\label{eq.input-output}
	\bv(\bk,\omega)
	\;=\;
	T_{\bv \bd}(\bk,\omega)\,
	\bd(\bk,\omega)
	\; = \;
	C(\bk)
	\left(
	\mri \omega I \,-\, A(\bk)
	\right)^{-1}
	B(\bk)
	\,
	\bd(\bk,\omega).
\end{align}
The steady-state output covariance matrix $V (\bk)$ is related to the spectral density matrix $S_{\bv \bv} (\bk, \omega)$ via,
\begin{align}
\label{eq.relation-V-S}
	V (\bk)
	\; \DefinedAs \;
	R_{\bv \bv}(\bk, 0)
	\;=\;
	\dfrac{1}{2 \pi}\,
	\ds{\int^{+\infty}_{-\infty}} S_{\bv \bv}(\bk, \omega) \, \mrd \omega.
\end{align}
Finally, for white-in-time input $\bd(\bk,t)$ in Equation~\ref{eq.lnse1}, the lagged output covariance matrix $R_{\bv \bv}(\bk, \tau)$ can be expressed as a linear function of the steady-state covariance matrix $X(\bk)$,
	\be
	\ba{rcl}
	R_{\bv \bv}(\bk,\tau)
	& = &
	C(\bk) X(\bk)\, \mre^{A^*(\bk) \tau} C^*(\bk).
	\ea
\ee

\begin{textbox}[h]
\section{ADMISSIBLE COVARIANCES}
The matrix $X = X^* \succ 0$ is the stationary covariance matrix of the state of the LTI system in Equation~\ref{eq.lnse1} with controllable pair $(A,B)$ and suitable input process $\bd$ if and only if
\begin{subequations}
\begin{align}
	\label{eq.RankConstraint-Sigma}
	\rank
	\left[
	\begin{matrix}
	AX \,+\, X A^*  & B
	\\
	B^* & 0
	\end{matrix}
	\right]
	\, = \;
	\rank \left[\begin{matrix}
	0 & B
	\\
	B^* & 0
	\end{matrix}\right],
\end{align}
or equivalently, if and only if the matrix equation
\begin{align}
	B  \,H^* 
	\,+\,
	H  \,B^* 
	\;=\;
	-\left(
	A  \, X
	\,+\, 
	X  \,A^* 
	\right),
	\label{eq.BH}
\end{align}
\end{subequations}
has a solution $H$~\cite{geo02b,geo02a}. The rank condition in Equation~\ref{eq.RankConstraint-Sigma} implies that any positive-definite matrix $X$ is admissible as a stationary covariance of the state of an LTI system if the input matrix $B$ is full row rank.
\end{textbox}

	\vspace*{-3ex}
\subsubsection*{Summary} For the LTI dynamics in Equation~\ref{eq.lnse1}, the algebraic constraint in Equation~\ref{eq.lyap_BH} determines admissible steady-state covariance matrices $X (\bk)$. Among all positive semi-definite matrices, this constraint identifies those that qualify as state-covariances for a state-space representation with matrices $A(\bk)$ and $B (\bk)$. As shown in~\cite{geo02b,geo02a}, the structure of state-covariances is an inherent property of the linear dynamics. The sidebar ADMISSIBLE COVARIANCES describes necessary and sufficient conditions for a positive-definite matrix $X(\bk)$ to qualify as a steady covariance matrix of the state $\bpsi (\bk,t)$ in Equation~\ref{eq.lnse1}. These conditions amount to the solvability of Equation~\ref{eq.BH} for the matrix $H (\bk)$ or, equivalently, the rank condition in Equation~\ref{eq.RankConstraint-Sigma}. We next build on such structural constraints on admissible covariances and formulate convex optimization problems for characterizing the statistical properties of stochastic excitations to LTI systems that account for partially available statistics in turbulent channel flow. 

	\vspace*{-2ex}
\section{COMPLETION OF PARTIALLY AVAILABLE FLOW STATISTICS}
\label{sec.CCP}

The algebraic relations described in Section~\ref{sec.alg-relations} can be used to compute the steady covariance matrix $X (\bk)$ of the stochastically-forced LTI system in Equation~\ref{eq.lnse1} based on the linearized model (i.e., the matrices $A (\bk)$ and $B (\bk)$) and the input statistics. In stochastic dynamic modeling of turbulent flows, however, the converse question is of interest: starting from the covariance matrix $X (\bk)$ and the dynamic matrix $A (\bk)$ in Equation~\ref{eq.lnse1}, the objective is to identify the directionality of the disturbance (i.e., the matrix $B (\bk)$ in Equation~\ref{eq.lnse1}) and the power spectrum of the stochastic input $\bd (\bk,t)$ that generate such state statistics. As illustrated in {\bf Figure~\ref{fig.filter}}, this amounts to designing a linear filter which is driven by white noise and produces input $\bd (\bk,t)$ that generates the desired covariance matrix $X (\bk)$ for the LTI system in Equation~\ref{eq.lnse1}. In high-Reynolds-number flows, experimental and computational limitations often lead to only partial knowledge of flow statistics. For example, in experiments, an array of probes may only provide a limited subset of spatio-temporal correlations for velocity fluctuations, and in numerical simulations, certain regions of the computational domain may be poorly resolved. In this section, we formulate the problem of completing partially known state correlations in a way that is consistent with the hypothesis that perturbations around the turbulent mean velocity are generated by the linearized NS equations. To accomplish this objective, we seek stochastic forcing models of low-complexity where complexity is quantified by the number of degrees of freedom that are directly influenced by stochastic forcing in the linearized evolution model.

\begin{figure}
	\begin{center}
	\begin{tabular}{c}
		\subfigure[]{
		         %_______________________________________________________________________________
%
%   tikz figure for inclusion in exercises:
%
%   clp_2dof_output_pert_config:  closed loop system with perturbed models.  This uses
%                                              and input perturbation and the 2-DOF structure used for
%                                              the satellite example.
%
%   Marcello Colombino, 4 May 2015
%  
%_______________________________________________________________________________
%
% TikZ styles for drawing
%
\input{figures/Tikz_common_styles}
%
%   set a filename for externalization
% \tikzsetnextfilename{clp_2dof_input_pert_config}
%
\noindent
\begin{tikzpicture}[scale=1, auto, >=stealth']
  
    \footnotesize

   % specify with respect to the plant
    
     \node[block, minimum height = .8cm, top color=Red!10, bottom color=Red!10] (sys1) {filter};
     
     \node[block, minimum height = 1.2cm, top color=RoyalBlue!10, bottom color=RoyalBlue!10] (sys2) at ($(sys1.east) + (2.8cm,0)$) {$\ba{c} \mbox{linearized} \\ \mbox{dynamics}\ea$};
     
     \node[] (output-node) at ($(sys2.east) + (2.2cm,0)$) {};
     
     \node[] (input-node) at ($(sys1.west) - (1.8cm,0)$) {}; 
      
%     \node[sum, anchor=west](usum) at ($(G.west) - (5.75cm,0)$) {+}; 
%     
%     \node[block, anchor=west] (delta) at ($(usum) + (0.5cm,0.75cm)$) {$\Delta(s)$};
%     
%     \node[block, anchor=east] (W) at ($(delta.east) + (2. 5cm,0cm)$) {$W_I(s)$}; 
%      
%     \node[block,  anchor=west ] (K) at ($(G.east) + (0.75cm,0)$) {$k$} ; 
%      
%     \node[branch] (ubranch) at ($(delta.east) + (3cm,-0.75cm)$) {};
%
%     \node [sum, anchor=west] (esum) at ($(ubranch) + (4.55cm,0)$) {$+$};
%      
%     \node[guide] (y) at ($(G.south) - (0,0.5cm)$){};
%
%     \node[guide] (R) at ($(esum.east) + (1,0.0cm)$){};

    % now link the nodes

    \draw [connector] (input-node) -- node [midway, above] {$\ba{c} \mbox{white} \\ \mbox{noise} \ea$} node [midway, below] {$\bw$} (sys1.west);
    \draw [connector] (sys2.east) -- node [midway, above] {$\ba{c} \mbox{velocity} \\ \mbox{fluctuations} \ea$} node [midway, below] {$\bv$} (output-node);
    \draw [connector] (sys1.east) -- node [midway, above] {$\ba{c} \mbox{colored} \\ \mbox{noise} \ea$} node [midway, below] {$\bd$} (sys2.west);
%    \draw [connector] (W) --   node [midway, above] {$z$} (delta);
%    \draw [connector] (ubranch) |-  (W);
%    \draw [connector] (delta) -| node [midway, above] {$v$}  (usum);
%    \draw [connector] (G) -- (usum); 
%    \draw [connector] (usum) -- node [near end, above, yshift=6pt] {$y$} 
%                                   ($(ybranch)-(0.75cm,0)$);
%    \draw [line] (ybranch) |- (y.east);
%    \draw [connector] (y) -| node [pos=0.9, right]{$-$} (esum);
%    \draw [connector] (R) -- node[midway, above] {$r$} (esum);                                            
\end{tikzpicture}
%_______________________________________________________________________________
		         \label{fig.filter}
		         }
		        \\
		\subfigure[]{
		         %_______________________________________________________________________________
%
%   tikz figure for inclusion in exercises:
%
%   clp_2dof_output_pert_config:  closed loop system with perturbed models.  This uses
%                                              and input perturbation and the 2-DOF structure used for
%                                              the satellite example.
%
%   Marcello Colombino, 4 May 2015
%  
%_______________________________________________________________________________
%
% TikZ styles for drawing
%
\input{figures/Tikz_common_styles}
%
%   set a filename for externalization
% \tikzsetnextfilename{clp_2dof_input_pert_config}
%
\noindent
\begin{tikzpicture}[scale=1, auto, >=stealth']
  
    \footnotesize

   % specify with respect to the plant
    
     \node[block, minimum height = 1.2cm, top color=Red!10, bottom color=Red!10] (sys) {$\ba{c} \mbox{modified} \\ \mbox{dynamics}\ea$};
     
     \node[] (output-node) at ($(sys.east) + (2.2cm,0)$) {};
     
     \node[] (input-node) at ($(sys.west) - (1.8cm,0)$) {}; 
      
%     \node[sum, anchor=west](usum) at ($(G.west) - (5.75cm,0)$) {+}; 
%     
%     \node[block, anchor=west] (delta) at ($(usum) + (0.5cm,0.75cm)$) {$\Delta(s)$};
%     
%     \node[block, anchor=east] (W) at ($(delta.east) + (2. 5cm,0cm)$) {$W_I(s)$}; 
%      
%     \node[block,  anchor=west ] (K) at ($(G.east) + (0.75cm,0)$) {$k$} ; 
%      
%     \node[branch] (ubranch) at ($(delta.east) + (3cm,-0.75cm)$) {};
%
%     \node [sum, anchor=west] (esum) at ($(ubranch) + (4.55cm,0)$) {$+$};
%      
%     \node[guide] (y) at ($(G.south) - (0,0.5cm)$){};
%
%     \node[guide] (R) at ($(esum.east) + (1,0.0cm)$){};

    % now link the nodes

    \draw [connector] (input-node) -- node [midway, above] {$\ba{c} \mbox{white} \\ \mbox{noise} \ea$} node [midway, below] {$\bw$} (sys.west);
    \draw [connector] (sys.east) -- node [midway, above] {$\ba{c} \mbox{velocity} \\ \mbox{fluctuations} \ea$} node [midway, below] {$\bv$} (output-node);
%    \draw [connector] (W) --   node [midway, above] {$z$} (delta);
%    \draw [connector] (ubranch) |-  (W);
%    \draw [connector] (delta) -| node [midway, above] {$v$}  (usum);
%    \draw [connector] (G) -- (usum); 
%    \draw [connector] (usum) -- node [near end, above, yshift=6pt] {$y$} 
%                                   ($(ybranch)-(0.75cm,0)$);
%    \draw [line] (ybranch) |- (y.east);
%    \draw [connector] (y) -| node [pos=0.9, right]{$-$} (esum);
%    \draw [connector] (R) -- node[midway, above] {$r$} (esum);                                            
\end{tikzpicture}
%_______________________________________________________________________________
		         \label{fig.sys-modified}
		         }
	\end{tabular}
	\end{center}
	\caption{(a) The cascade connection of the linearized dynamics with a spatio-temporal linear filter which is designed to account for partially available second-order statistics of turbulent channel flow; (b) An equivalent reduced-order representation of the cascade connection in (a).}
	\label{fig.filter_sys-feedback}
\end{figure}

		\vspace*{-2ex}
\subsection{Necessity for colored-in-time stochastic forcing}

\begin{figure}
	\hspace{2.8cm}
	\begin{tabular}{rl}
	\begin{tabular}{c}
	\vspace{.6cm}
	\normalsize{\rotatebox{90}{$\lambda_i \left(A \, X_{\mathrm{dns}} + X_{\mathrm{dns}} \, A^*\right)$}}
	\end{tabular}
	&
	\hspace{-4.4cm}
	\begin{tabular}{c}
        \includegraphics[width=6cm]{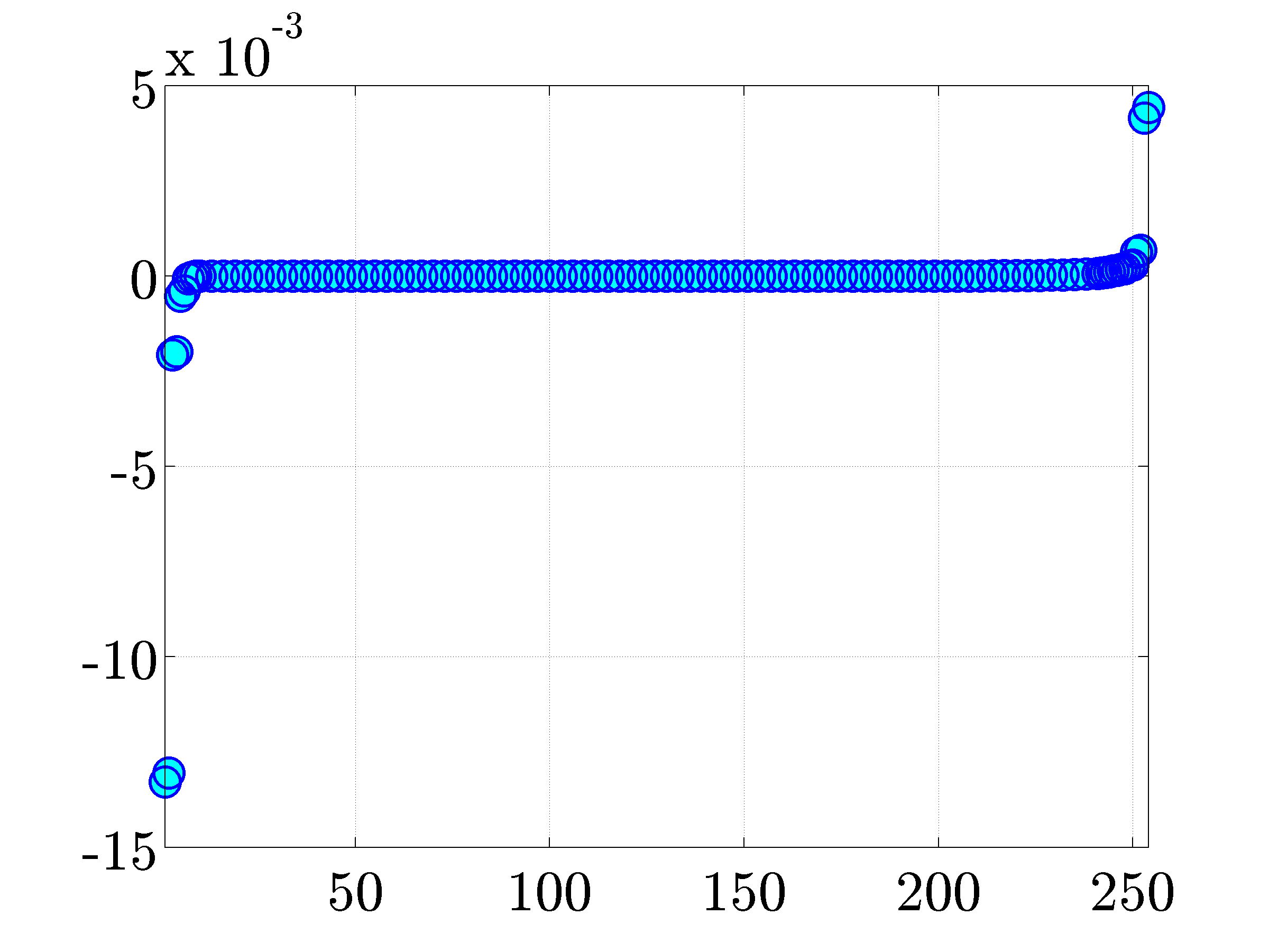}
        \\[-.1cm]
        \normalsize{$i$}
        \end{tabular}
        \end{tabular} 
\caption{Positive eigenvalues of the matrix $A (\bk) X_{\mathrm{dns}} (\bk) + X_{\mathrm{dns}} (\bk) A^* (\bk)$ for channel flow with $Re=186$ and $\bk = (2.5,7)$, indicate that turbulent velocity covariances cannot be reproduced by the linearized NS equations with white-in-time stochastic forcing (cf.\ Equation~\ref{eq.standard_lyap}).}
\label{fig.Forcing_evalues}
\end{figure}

The right-hand-side of standard algebraic Lyapunov equation~\ref{eq.standard_lyap} is sign-definite, i.e., $B(\bk)\, \Omega(\bk) B^*(\bk) \succeq 0$. In contrast, the right-hand-side of Lyapunov-like equation~\ref{eq.lyap_BH} is in general sign-indefinite and, unless the input $\bd (\bk,t)$ in Equation~\ref{eq.lnse1} is white-in-time, matrix 
\begin{align}
	\label{eq.rhs_lyap}
	\ba{rcl}
	Z (\bk)
	& \DefinedAs &
	- \left(A(\bk)\,X(\bk)
	\,+\,
	X(\bk)\, A^*(\bk)
	\right)
	\; = \;
	B(\bk)\, H^*(\bk)
	\,+\,
	H(\bk)\, B^*(\bk),
	\ea
\end{align}
can have both positive and negative eigenvalues. {\bf Figure~\ref{fig.Forcing_evalues}} shows the eigenvalues of the matrix $A (\bk) X_{\mathrm{dns}} (\bk)+ X_{\mathrm{dns}} (\bk) A^* (\bk)$ for a channel flow with $Re=186$ and $\bk = (2.5,7)$, where $A (\bk)$ denotes the generator of the dynamics in Equation~\ref{eq.lnse1} obtained by linearization around the turbulent mean velocity profile and $X_{\mathrm{dns}} (\bk)$ is the steady-state covariance matrix resulting from numerical simulations of the nonlinear NS equations. The presence of both positive and negative eigenvalues indicates that the second-order statistics of turbulent channel flow cannot be reproduced by the linearized NS equations with white-in-time stochastic excitation. The modeling and optimization framework that was recently developed in~\cite{zarchejovgeoTAC17,zarjovgeoJFM17} overcomes this limitation by departing from the white-in-time restriction on \mbox{stochastic~forcing.}
	
		\vspace*{-2ex}
\subsection{Covariance completion via convex optimization}
\label{sec.cc-noisemodeling}

\begin{figure}
\begin{center}
         \includegraphics[width=3.4cm]{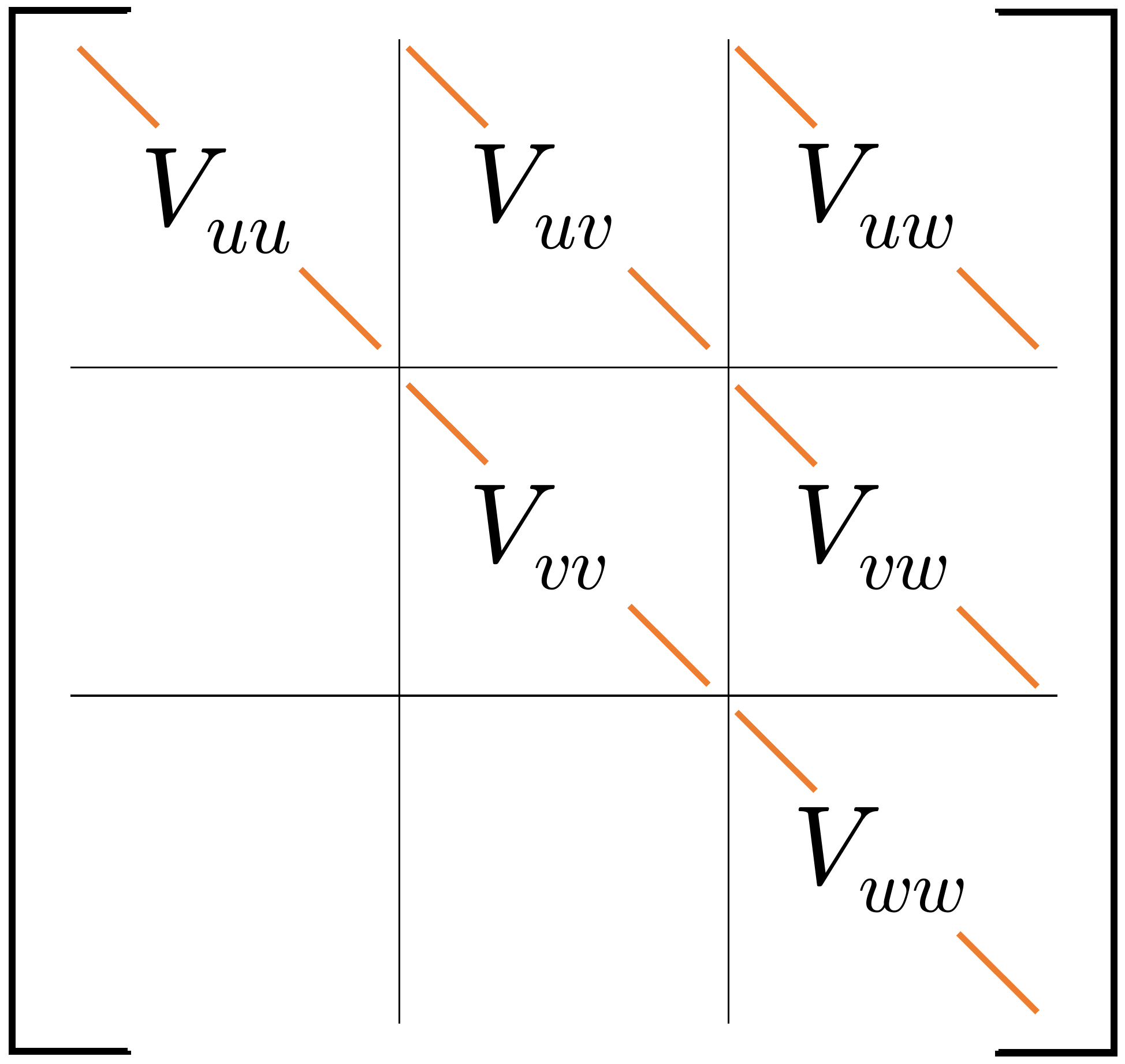}
\end{center}
\caption{Structure of the output covariance matrix $V$. Available one-point correlations of the velocity vector $\bv$ in the wall-normal direction at various wavenumbers $\bk$ are marked by the orange lines.}
\label{fig.output_covariance}
\end{figure}

For the dynamical generator $A$ resulting from linearization of the NS equations around the turbulent mean velocity, the steady-state covariance matrix $X$ satisfies
\begin{subequations}
\begin{align}
 	\label{eq.constraint_lyap}
 	A \, X 
	\;+\; 
	X A^* 
	\;+\; 
	Z  
	\;=\; 
	0,
\end{align}
where
	\begin{align}
	\label{eq.rhs_lyap1}
	Z 
	\; \DefinedAs \;
	B \, H^*
	\; + \;
	H \, B^*,
\end{align}
quantifies the contribution of stochastic excitation. For notational convenience, we omit the dependence on the wavenumber $\bk$ in this section. We assume that a subset of entries of the output covariance matrix $V$, namely $V_{ij}$ for a selection of indices $(i,j)\in \mathcal I$, is available. This yields an additional set of linear constraints for the matrix $X$,
\begin{align}
	\label{eq.constraint_observations}
	(CXC^*)_{ij}
	\;=\;
	V_{ij}, ~~ (i,j)\in \mathcal I.
\end{align}
\end{subequations}
For example, such known entries may represent one-point correlations in the wall-normal direction; see {\bf Figure~\ref{fig.output_covariance}} for an illustration. At any $\bk$, the diagonals of the submatrices $V_{uu} (\bk)$, $V_{vv} (\bk)$, and $V_{ww} (\bk)$ denote the normal Reynolds stresses \mbox{in turbulent channel flow, e.g.,} 
	\[
	\diag \left(V_{uu} (\bk) \right) 
	\; = \;
	\diag
	\left( 
	\lim_{t \, \to \, \infty}
	\bE
	\, 
	( 
	u (\bk,t)\, u^*(\bk,t) 
	)
 \right),
	\]
and the main diagonal of the submatrices $V_{uv} (\bk)$, $V_{uw} (\bk)$, and $V_{vw} (\bk)$ denote the shear stresses, e.g., 
	$
	\diag \left(V_{uv} (\bk) \right) 
	= 
	\diag
	\,
	( 
	\lim_{t \, \to \, \infty}
	\bE
	\, 
	( 
	u (\bk,t)\, v^*(\bk,t) 
	)).
	$
It is noteworthy that while the covariance matrix $X$ is not allowed to have negative eigenvalues, the matrix $Z$ can be sign indefinite. Our objective is to identify suitable choices of $X$ and $Z$ that satisfy the above constraints and yield a low-complexity model for the stochastic input that explains the observed statistics. 

\begin{marginnote}[]
In statistical steady-state, turbulent kinetic energy is determined by the sum of traces of matrices $V_{uu}$, $V_{vv}$, and $V_{ww}$ and skin-friction drag depends on the shear~stress $\diag \left( V_{uv} \right)$.
\end{marginnote}

The contribution of the stochastic excitation enters through the matrix $Z$, which is of the form given by Equation~\ref{eq.rhs_lyap1}, where the directionality of the input and its time-correlations are reflected by the choices of matrices $B$ and $H$. As discussed in the sidebar ADMISSIBLE COVARIANCES, when the input matrix $B$ is full rank any positive semi-definite $X$ qualifies as the steady-state covariance of stochastically-forced linearized NS equations~\ref{eq.lnse1}. However, as demonstrated in~\cite{zarjovgeoJFM17}, in this case a forcing model that cancels the linearized dynamics and obscures important aspects of the underlying physics becomes a viable option. It is thus important to minimize the number of degrees of freedom that can be directly influenced by stochastic forcing. This can be accomplished via suitable regularization, e.g., by minimizing rank of the matrix $Z$ in Equation~\ref{eq.constraint_lyap}~\cite{chejovgeoCDC13,zarchejovgeoTAC17}.
\begin{marginnote}[]
The rank of the matrix $Z$ bounds the number of independent input channels or columns in matrix $B$; see~\cite{zarchejovgeoTAC17} for details.
\end{marginnote}

This gives rise to the convex optimization problem~\ref{eq.CP1} described in the sidebar COVARIANCE COMPLETION PROBLEM.  The objective function in~\ref{eq.CP1} provides a trade-off between the solution to the maximum entropy problem and the complexity of the forcing model, where the positive regularization parameter $\gamma$ reflects the relative weight of the nuclear norm objective. The convexity of optimization problem~\ref{eq.CP1} follows from the convexity of the objective function and the linearity of the constraint set, thereby implying the existence of a globally optimal minimizer.

%\begin{marginnote}[]
%\tc{red}{Comment on history of CC problems.}
%\end{marginnote}

\begin{textbox}[t]
\section{COVARIANCE COMPLETION PROBLEM}
Given matrices $A$ and $C$, the available entries $V_{ij}$ of the output covariance matrix $V$, and positive regularization parameter $\gamma$, determine the Hermitian matrices $X$ and $Z$ that solve convex optimization problem
	\begin{align} 
	\ba{cl}
	\minimize\limits_{X, \, Z}
	& 
	-\logdet\left(X\right) 
	\; + \; 
	\gamma \, \norm{Z}_\star
	\\[.15cm]
	\subject 
	&
	~A \, X \,+\, X A^* \,+\, Z  \;=\; 0
	\\[0.05cm]
	&
	\, (CXC^*)_{ij} \;=\;V_{ij}, ~ (i,j)\in \mathcal I.
	 \ea
	 \tag{CC-1}
	\label{eq.CP1}
\end{align} 
The first constraint reflects the requirement that the second-order statistics are consistent with stochastically-forced linearized model and the second constraint requires that the available elements of the matrix $V$ are exactly reproduced. Minimizing the logarithmic barrier function ensures positive-definiteness of the covariance matrix $X$~\cite{boyvan04} and results in a maximum entropy stochastic realization~\cite{goopay77}. On the other hand, the nuclear norm regularizer, i.e., the sum of singular values of the matrix $Z$, $\norm{Z}_\star = \sum_i \sigma_i (Z)$, is introduced to restrict the rank of $Z$~\cite{faz02,recfazpar10} and thereby reduce the complexity of the forcing model. 
\end{textbox}

	\vspace*{-2ex}
\subsubsection{Power spectrum of stochastic input and filter design}
\label{sec.filter}

The solution of problem~\ref{eq.CP1}, i.e., the Hermitian matrices $X (\bk)$ and $Z (\bk)$, can be used to obtain a dynamical model for colored-in-time stochastic input to the linearized NS equations. A class of generically minimal linear filters, which have the same number of degrees of freedom as the finite-dimensional approximation of the linearized model, was recently developed in~\cite{zarchejovgeoTAC17,zarjovgeoJFM17}.

Since channel flow is translationally invariant in the wall-parallel dimensions, the dynamics in Equation~\ref{eq.lnse1} and optimization problem~\ref{eq.CP1} are decoupled over the wavenumbers $\bk = (k_x,k_z)$. At each $\bk$, the filter dynamics that account for $X (\bk)$ are given by
\be
\label{eq.filter}
\ba{rcl}
	   \dot{\bphi} (\bk,t)
	   & = &
	    (A (\bk) \,-\, B (\bk) K (\bk)) 
	    \,
	    \bphi (\bk,t)
	    \; + \;
	    B (\bk)
	    \,
	    \bw (\bk,t),
	    \\[0.15cm]
	    \bd (\bk,t)
	    & = &
	    - K (\bk)
	    \,
	    \bphi (\bk,t)
	    \; + \;
	    \bw (\bk,t),
\ea
\ee
where $\bphi (\bk,t)$ is the state of the filter and $\bw (\bk,t)$ is a zero-mean white-in-time stochastic process with covariance $\Omega (\bk)$. On the other hand,
\begin{align}
\label{eq.filter-C}
	K (\bk)
	\;=\;
	(\tfrac{1}{2} \, \Omega (\bk) B^* (\bk) -\, H^* (\bk)) \, X^{-1} (\bk),
\end{align}
for matrices $B  (\bk)$ and $H (\bk)$ that correspond to the factorization $Z (\bk) = B (\bk) H^* (\bk) + H (\bk) B^* (\bk)$; see~\cite{zarchejovgeoTAC17} for details. The linear filter represented by Equation~\ref{eq.filter} generates a {\em colored-in-time\/} stochastic input $\bd (\bk,t)$ to the linearized NS equations~\ref{eq.lnse1} and the resulting cascade connection reproduces the available second-order statistics of turbulent flow. The spectral density of $\bd (\bk,t)$
\begin{align*}
%	\label{eq.power-spectrum}
	S_{\bd \bd} (\bk,\omega)
	\; = \; 
	T_{\bd \bw} (\bk,\omega)
	\,
	\Omega (\bk)
	\,
	T_{\bd \bw}^* (\bk,\omega),
\end{align*}
determines the spectral content of the input to the LTI system, where 
	\be
	T_{\bd \bw} (\bk,\omega) 
	\; = \;
	K (\bk) \left( \mri \omega I \, - \, A (\bk) \, + \, B (\bk) K (\bk) \right)^{-1} B (\bk) \; + \; I,
	\non
	\ee
is the spatio-temporal frequency response of the linear filter in Equation~\ref{eq.filter}.

		\vspace*{-2ex}
\subsubsection{Minimal realization}
The state-space representation corresponding to the cascade connection of the linear filter in Equation~\ref{eq.filter} with the linearized NS dynamics in Equation~\ref{eq.lnse1} is given by
\begin{align}
 	\label{eq.cascade}
 	\ba{rcl}
             	\tbo{\dot{\bpsi} (\bk,t)}{\dot{\bphi} (\bk,t)}
            	& \;=\; &
            	\tbt{A (\bk)~}{~-B (\bk) K (\bk)}{0~}{~A (\bk) - B (\bk) K (\bk)}
            	\tbo{\bpsi (\bk,t)}{\bphi (\bk,t)}
            	\, + \,
            	\tbo{B (\bk)}{B (\bk)}
		\bw (\bk,t)
		\\[0.2cm]
		\bv (\bk,t)
		& \;=\; &
		\obt{C (\bk)~}{~0}
		\tbo{\bpsi (\bk,t)}{\bphi (\bk,t)}.
	\ea
 \end{align}
This realization has twice as many states as the spatial discretization of the linearized NS model in Equation~\ref{eq.lnse1}, but is not controllable and therefore not minimal. As shown in~\cite{zarchejovgeoTAC17}, removal of the uncontrollable states yields the minimal realization of the mapping from the input $\bw (\bk,\omega)$ to the output $\bv (\bk,\omega)$, $\bv (\bk,\omega) = T_{\bv \bw} (\bk,\omega) \bw (\bk,\omega)$,
	\begin{align}
	T_{\bv \bw} (\bk,\omega) 
	\; = \;
	C (\bk) \left( \mri \omega I \, - \, A (\bk) \, + \, B (\bk) K (\bk) \right)^{-1} \! B (\bk),
	\label{eq.Tvw}
\end{align}
as
\begin{align}
	\label{eq.feedback_dyn}
	\ba{rcl}
        \dot{\bpsi} (\bk,t)
        & \; = \; &
        \left( A (\bk) \, - \, B (\bk) K (\bk)\right) 
        \bpsi (\bk,t)
        \;+\; 
        B (\bk) \, \bw (\bk,t),
        \\[0.1cm]
        \bv (\bk,t)
	& \; = \; &
	C (\bk) \,
	\bpsi (\bk,t).
	\ea
\end{align}
This system has the same number of degrees of freedom as the system in Equation~\ref{eq.lnse1} and the corresponding algebraic Lyapunov equation in conjunction with Equation~\ref{eq.filter-C} yield
\begin{align*}
	\ba{l}
	\left( A (\bk) \, - \, B (\bk) K(\bk) \right) X (\bk)
	\,+\,
	X (\bk) \left( A (\bk) \, - \, B (\bk) K (\bk) \right) ^*
	\,+\,
	B (\bk) \,\Omega (\bk) \, B^*(\bk)
	\\[.1cm]
	\hspace{0.25cm}
	=\;
	A (\bk) \,X (\bk) \,+\, X (\bk) \,A^* (\bk) \,+\, B (\bk) \, \Omega (\bk) \, B^* (\bk) 
	\,-\, B (\bk) K (\bk) X (\bk)  \,-\, X (\bk) K^* (\bk) B^* (\bk)
	\\[.1cm]
	\hspace{0.25cm}
	=\;
	A (\bk) \,X (\bk) \,+\, X (\bk) \,A^* (\bk) \,+\, B (\bk) \,H^* (\bk) \,+\, H (\bk) \, B^* (\bk)
	\; = \; 0.
	\ea
\end{align*}
This demonstrates that the state-space realization in Equation~\ref{eq.filter} generates a stochastic input $\bd (\bk,t)$ which is consistent with the steady-state covariance matrix $X (\bk)$. 

\begin{marginnote}[]
The effect of colored-in-time excitation is equivalent to white-in-time excitation together with a structural perturbation of the system dynamics.
\end{marginnote}

	\vspace*{-1ex}
\begin{remark}
From Equation~\ref{eq.filter-C} we have
	$
	H (\bk) 
	= 
	\tfrac{1}{2} B (\bk) \Omega (\bk) -  X (\bk) K^* (\bk)
	$
and substitution of this expression into Equation~\ref{eq.lyap_BH} yields the standard algebraic Lyapunov equation
    \begin{align*}
    	(A (\bk) \,-\,B (\bk) K (\bk) ) X (\bk) 
	\;+\; 
	X (\bk) (A (\bk) \,-\, B (\bk) K (\bk) )^*
	\;=\;
	-B (\bk) \, \Omega (\bk) B^* (\bk).
    \end{align*}
Since the pair $(A (\bk), B (\bk))$ is controllable, so are $(A (\bk) - B (\bk) K (\bk), B (\bk))$ and $(A (\bk) - B (\bk) K (\bk), B (\bk) \, \Omega^{1/2} (\bk))$. Stability of the modified dynamical generator $A (\bk) - B (\bk) K (\bk)$ follows from positive semi-definiteness of $B (\bk) \, \Omega (\bk) B^* (\bk)$ via standard Lyapunov theory.
	\label{rem.stability}
	\end{remark}
	\vspace*{-1ex}
	
The minimal realization (given by Equation~\ref{eq.feedback_dyn}) of the cascade connection described by Equation~\ref{eq.cascade} is advantageous from a computational standpoint and it allows for an alternative interpretation of the stochastic realization of colored-in-time forcing. First, time-domain simulations require numerical integration of the system in Equation~\ref{eq.feedback_dyn}, which has half the number of states as compared to the system in Equation~\ref{eq.cascade}, thereby offering computational speedup. On the other hand, the structure in Equation~\ref{eq.feedback_dyn} suggests that the colored-in-time forcing realized by the LTI filter in Equation~\ref{eq.filter} can be equivalently interpreted as a dynamical modification to the linearized equations in the form of state-feedback interactions. This interpretation provides an alternative viewpoint that is closely related to a class stochastic control~\cite{HotSke87,yasskegri93,grikarske94,chegeopav16b} and output covariance estimation~\cite{linjovTAC09,zorfer12} problems; see~\cite[Section II.C]{zarchejovgeoTAC17} for details. Based on this, we next describe an alternative formulation of the covariance completion problem as a state-feedback synthesis that is optimal with respect to a different design criterion~\cite{zarjovgeoCDC16,zarmohdhijovgeoTAC18}.

		\vspace*{-2ex}
\subsection{Minimum-control-energy covariance completion problem}
\label{sec.ProxKCC}

As described in~\cite{zarjovgeoCDC16,zarmohdhijovgeoTAC18}, the challenge of establishing consistency between statistical measurements and a linearized model can be alternatively cast as the problem of seeking a completion of the missing entries of a covariance matrix $X$ along with a perturbation $\Delta$ of the system dynamics subject to white-in-time input $\bw$,
\begin{align*}
        \ba{rcl}
            \dot{\bpsi}
            & \; = \; &
            (A \,+\, \Delta)
            \,
            \bpsi
            \; + \;
            \bw,
            \\[0.05cm]
            \bv
            & \; = \; &
            C
            \,
            \bpsi.
        \ea
\end{align*}
For $\Delta \DefinedAs -BK$, a covariance completion problem can be formulated as an optimal control problem aimed at designing a stabilizing state-feedback control law ${\bf f} = -K \bpsi$ ({\bf Figure~\ref{fig.sysfeedback}}). The choice of $B$ may incorporate added insights into the strength and directionality of possible couplings between state variables. While a full-rank matrix $B$ that allows the perturbation signal $K \bpsi$ to manipulate all degrees of freedom can lead to the complete cancellation of the original dynamics $A$, it is also important to impose a penalty on the average quadratic size of signals $K \bpsi$. This gives rise to convex optimization problem~\ref{eq.CP2} described in the sidebar MINIMUM ENERGY COVARIANCE COMPLETION PROBLEM. The objective function in~\ref{eq.CP2} provides a trade-off between the minimum-control-energy problem and the number of feedback couplings that need to be introduced to modify the dynamical generator $A$ and achieve consistency with available data~\cite{zarjovgeoCDC16,zarmohdhijovgeoTAC18}.
%\begin{marginnote}[]
%\tc{red}{Comment on history of minimum energy control problems.}
%\end{marginnote}

\begin{textbox}[t]
\section{MINIMUM ENERGY COVARIANCE COMPLETION PROBLEM}
Given matrices $A$, $B$, $C$, $R$, $\Omega$, the available entries $V_{ij}$ of the output covariance matrix $V$, and the positive regularization parameter $\gamma$, determine the matrices $K$ and $X$ that solve convex optimization problem
	\begin{align} 
	\ba{rl}
	\minimize\limits_{K,\, X}
	& 
	~
	\trace \left( K^*R\,K X \right)
	\; + \; 
	\gamma \, \ds{\sum^n_{i \, = \, 1} w_i \norm{\mre_i^* K}_2}
	\\[.15cm]
	\subject 
	&
 	~
	(A - BK) \, X \,+\, X (A - BK)^* \,+\, \Omega  \;=\; 0
	\\[0.05cm]
	&
	~
	(CXC^*)_{ij} \;=\;V_{ij}, ~~ (i,j) \, \in \, {\mathcal I}
	\\[0.05cm]
	&
	~
	X \, \succ\, 0.
	 \ea
	 \tag{CC-2}
	\label{eq.CP2}
\end{align}
The algebraic constraint on $K$ and $X$ ensures closed-loop stability (see {\bf Remark~\ref{rem.stability}}) and consistency with the state covariance matrix $X$, and the second equality constraint requires that the available elements of the matrix $V$ are exactly reproduced. The positive-definite matrix $R$ specifies a penalty on the control input while the weighted-norm regularizer promotes sparsity on the rows of the matrix $K$. Here, $w_i$ are given positive weights, $\mre_i$ is the $i$th unit vector in $\bbR^m$, and $\Omega \succeq 0$ is the covariance matrix of white noise input $\bw$.
\end{textbox}

\begin{figure}
\begin{center}
         %_______________________________________________________________________________
%
%   tikz figure for block-diagrams in presentations and papers:
%
%   Armin Zare, 16 May 2016
%  
%_______________________________________________________________________________
%
% TikZ styles for drawing
%
\input{figures/Tikz_common_styles}
\noindent
\begin{tikzpicture}[scale=1, auto, >=stealth']

 \footnotesize
    
     \node[block, minimum height = 1.2cm, top color=RoyalBlue!10, bottom color=RoyalBlue!10] (sys1) {\begin{tabular}{c}linearized\\[-.05cm] dynamics\end{tabular}};
     
     \node[block, minimum height = .8cm, top color=Red!10, bottom color=Red!10] (sys2) at ($(sys1.south) - (0cm,.8cm)$) {$-K$};
                          
     \node[] (R1) at ($(sys1.east) + (1.5cm,-0.3cm)$){};
     
%      \node[] (R2) at ($(sys1.west) - (3cm,0.7cm)$){};
%      
%      \node[] (R3) at ($(sys1.east) + (3cm,-0.7cm)$){};
     
    % now link the nodes
    
    \draw [connector] ($(sys1.west) + (-2.1cm,.3cm)$) -- node [midway, above] {$\ba{c} \mbox{white} \\ \mbox{noise} \ea$} node [midway, below] {$\bw$} ($(sys1.west) + (0cm,.3cm)$);
                    	
%    \draw [line] (sys1.east) -- (R1);
    
    \draw [line] (sys2.west) -|  ($(sys1.west) + (-1.5cm,-.3cm)$);
    
    \draw [connector] ($(sys1.west) + (-1.5cm,-.3cm)$) -- node [midway, below] {$\bf f$} ($(sys1.west) + (0cm,-.3cm)$);
    
    \draw [connector] ($(sys1.east) + (0cm,.3cm)$) -- node [midway, above] {$\ba{c} \mbox{velocity} \\ \mbox{fluctuations} \ea$} node [midway, below] {$\bv$} ($(sys1.east) + (2.1cm,.3cm)$);
    
    \draw [line] ($(sys1.east) + (0cm,-.3cm)$) -- node[midway, below] {$\bpsi$} (R1.west);
    
    \draw [connector] (R1.west) |- (sys2.east);
    
%    \draw [dashed] (R2) -- (R3);
                                       
\end{tikzpicture}
%_______________________________________________________________________________
\end{center}
\caption{A feedback connection of the linearized dynamics with a static gain matrix $K$ that is designed to account for the sampled steady-state covariance matrix~$X$.}
\label{fig.sysfeedback}
\end{figure}

	\vspace*{-1ex}
\begin{remark}
As demonstrated in~\cite{zarjovgeoACC15,zarchejovgeoTAC17,zarmohdhijovgeoTAC18}, covariance completion problems~\ref{eq.CP1} and~\ref{eq.CP2} can be cast as semidefinite programs. For small- and medium-size problems, these can be solved efficiently using standard solvers~\cite{SDPT3,cvx,boyvan04}. To deal with large problem dimensions that arise in fluid dynamics, customized algorithms have been developed in~\cite{zarchejovgeoTAC17,zarmohdhijovgeoTAC18}.
\end{remark}

	% \vspace*{-3ex}
\subsection{Completion of spatio-temporal correlations}

The covariance matrix $V(\bk)$ provides information about spatial correlations of velocity fluctuations in statistical steady-state. As described in Section~\ref{sec.spatio-temporal-corr}, the temporal dependence of such statistics is captured by the spectral density matrix $S_{\bv\bv}(\bk, \omega)$. This matrix can be used to provide real-time estimates of the flow state~\cite{saspiacavjor17}, and recent efforts have been directed at estimating $S_{\bv\bv}(\bk, \omega)$ by either matching individual entries at specified temporal frequencies~\cite{bensiparndanles16,benyegsiplec17,towlozyan19} or the spectral power~\cite{morsemhencos19}, $\trace(S_{\bv \bv}(\bk, \omega))$. Either way it should be independently considered whether the so-constructed colored-in-time forcing models preserve important aspects of the original linearized NS dynamics. For additional discussion on parsimonious models and how these may reflect underlying physics see Section~\ref{sec.cc-noisemodeling}.

	\vspace*{-2ex}
\section{CASE STUDY: TURBULENCE MODELING IN CHANNEL FLOW}
\label{sec.channel}

In this section, we investigate the completion of partially known second-order statistics of a turbulent channel flow using the framework presented in Section~\ref{sec.cc-noisemodeling}. The mean velocity profile and one-point velocity correlations in the wall-normal direction at various wavenumber pairs $\bk$ are obtained from DNS of a turbulent channel flow with friction Reynolds number $Re = 186$~\cite{kimmoimos87,moskimman99,deljim03,deljimzanmos04}; see {\bf Figures~\ref{fig.meanflow} and~\ref{fig.output_covariance}} for an illustration. We also show how the modified dynamics of Section~\ref{sec.filter} can be used as a low-dimensional model that is simulated in time to generate velocity fluctuations whose second-order statistics are consistent with numerical simulations of the nonlinear NS equations.

	\vspace*{-2ex}
\subsection{Reproducing available and completing unavailable second-order statistics}

\begin{figure}
%\centering
	\begin{tabular}{rcrc}
		\subfigure[]{\label{fig.Averaged_uu}}
		&&
		\subfigure[]{\label{fig.Averaged_uv}}
		\\[-.8cm]
		&
		\hspace{-3.9cm}
		\begin{tabular}{c}
			\includegraphics[width=6cm]{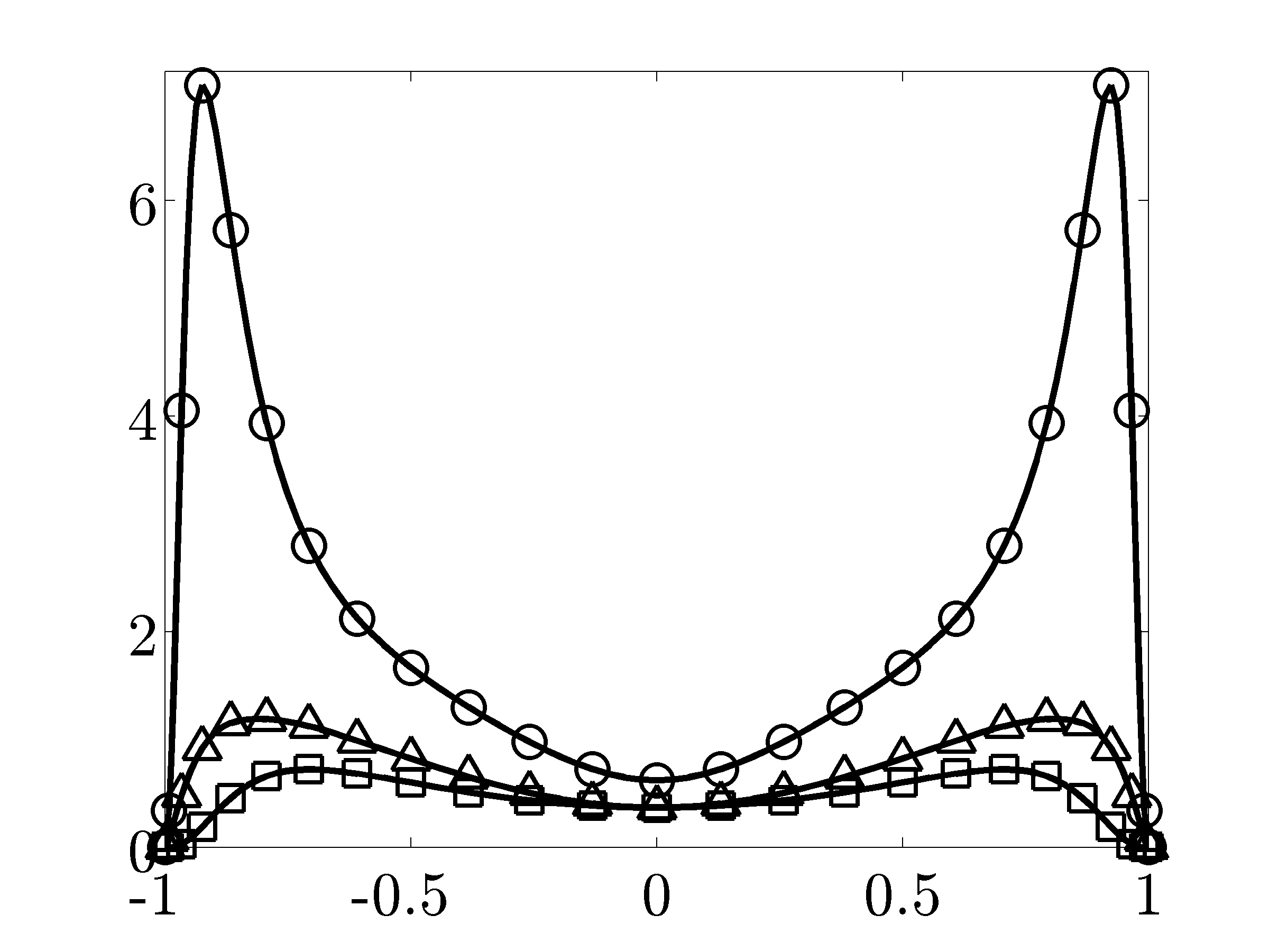}
			\\[-.2cm]
			\hspace{.1cm}
			{\normalsize $y$}
		\end{tabular}\hspace{-4cm}
		&&
		\hspace{-3.9cm}
		\begin{tabular}{c}
			\includegraphics[width=6cm]{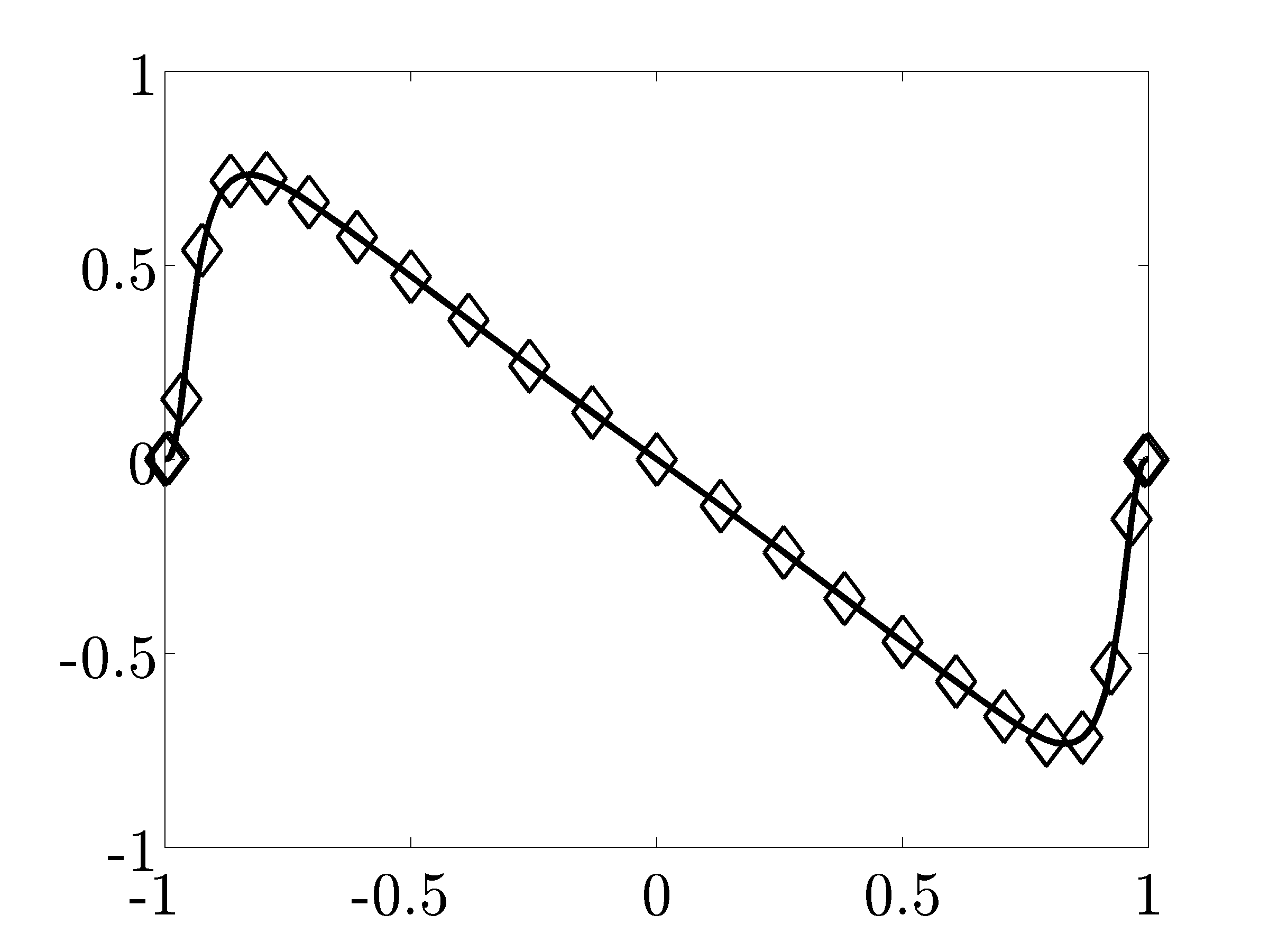}
			\\[-.2cm]
			\hspace{.1cm}
			{\normalsize $y$}
		\end{tabular}
	\end{tabular}
	\caption{(a) Correlation profiles of normal and (b) shear stresses resulting from DNS of turbulent channel flow with $Re=186$ (--) and from the solution to~\ref{eq.CP1}; $\diag \left(V_{uu} \right)$ ($\Circle$), $\diag \left(V_{vv} \right)$ ($\Box$), $\diag \left(V_{ww} \right)$ ($\triangle$), $-\diag \left(V_{uv} \right)$ ($\Diamond$). Profiles are integrated over the wall-parallel wavenumbers $\bk$.
	}
	\label{fig.averaged_profiles}
\end{figure}

As demonstrated in~\cite{zarjovgeoJFM17}, optimization problem~\ref{eq.CP1} is feasible at all wavenumbers $\bk$. Thus, regardless of the value of the regularization parameter $\gamma$, all available one-point correlations of turbulent channel flow can be reproduced by a stochastically-forced linearized model. {\bf Figure~\ref{fig.averaged_profiles}} displays perfect matching of all one-point velocity correlations that result from integration over wall-parallel wavenumbers. Since problem~\ref{eq.CP1} is not feasible for $Z (\bk) \succeq 0$ at all $\bk$, this {\em cannot\/} be achieved with white-in-time stochastic forcing. 

\begin{figure}
	\begin{tabular}{crccrc}
	\subfigure[]{\label{fig.Ruu_DNS}}
	&&&
	\subfigure[]{\label{fig.Ruu_NM}}
	&&
	\\[-.5cm]
	&
	\hspace{-2cm}
	\begin{tabular}{c}
		\vspace{.2cm}
		\normalsize{\rotatebox{90}{$y$}}
	\end{tabular}
	&
	\hspace{-4.3cm}
	\begin{tabular}{c}
		\normalsize{$V_{uu, \mathrm{dns}}$} 
		\\
		\includegraphics[width=6cm]{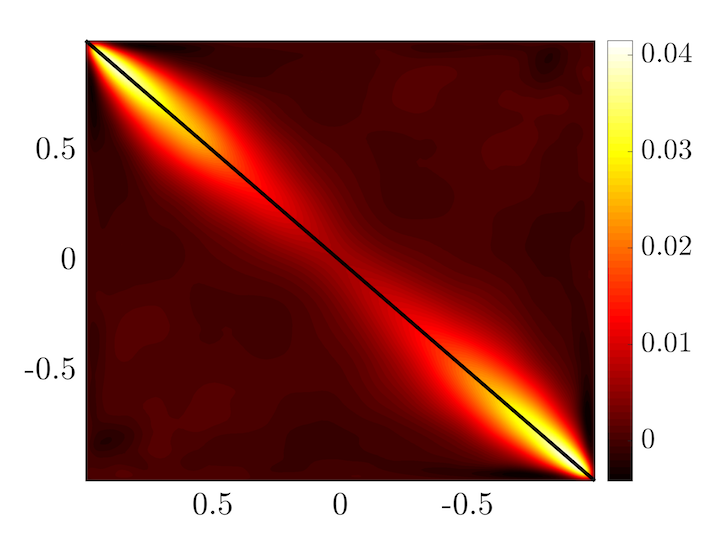}
	\end{tabular}
	\hspace{-4.15cm}
	&&&
	\hspace{-4.6cm}
	\begin{tabular}{c}
		\normalsize{$V_{uu}$}
		\\
		\includegraphics[width=6cm]{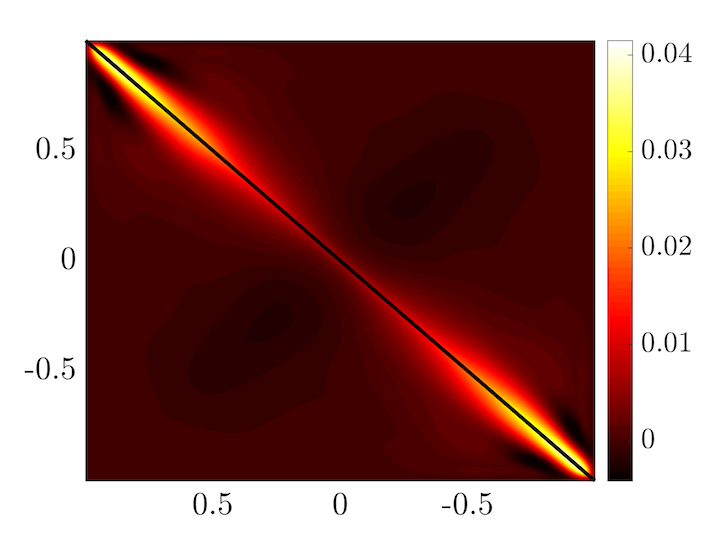}
	\end{tabular}
	\\[-.2cm]
	\subfigure[]{\label{fig.Ruv_DNS}}
	&&&
	\subfigure[]{\label{fig.Ruv_NM}}
	&&
	\\[-.5cm]
	&
	\hspace{-2cm}
	\begin{tabular}{c}
		\vspace{.4cm}
		\normalsize{\rotatebox{90}{$y$}}
	\end{tabular}
	&
	\hspace{-4.3cm}
	\begin{tabular}{c}
		\normalsize{$V_{uv, \mathrm{dns}}$} 
		\\
		\includegraphics[width=6cm]{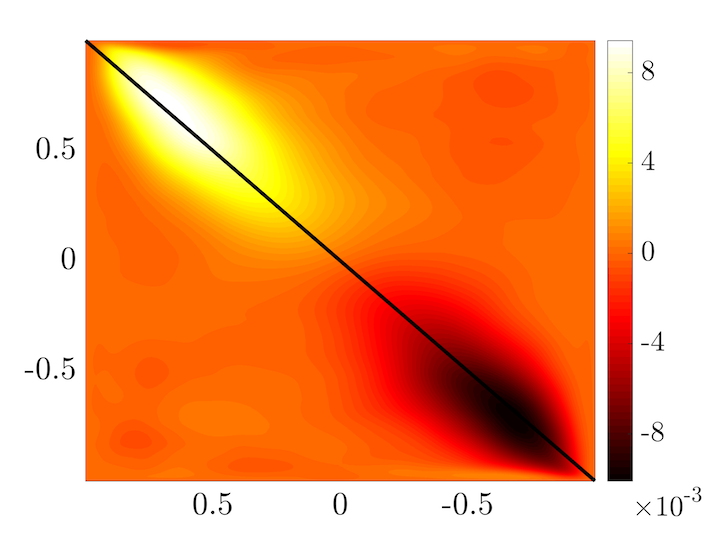}
		\\[-.1cm]
		\hspace{-.5cm}
		\normalsize{$y$}
	\end{tabular}
	\hspace{-4.15cm}
	&&&
	\hspace{-4.6cm}
	\begin{tabular}{c}
		\normalsize{$V_{uv}$}
		\\
		\includegraphics[width=6cm]{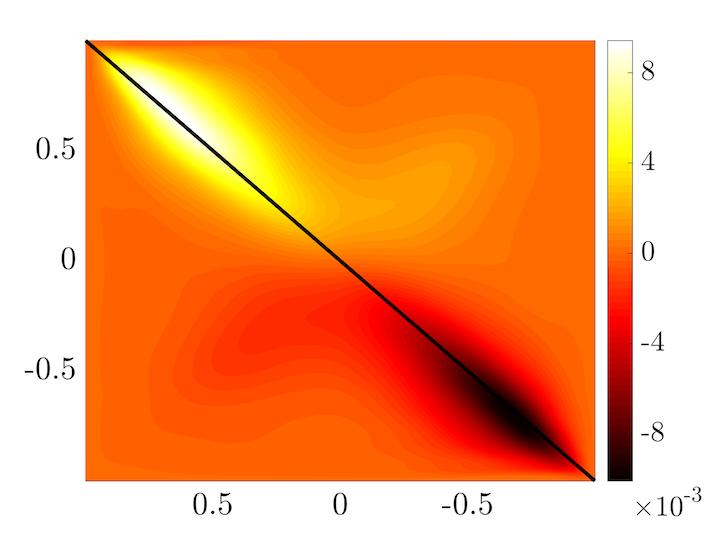}
		\\[-.1cm]
		\hspace{-.5cm}
		\normalsize{$y$}
	\end{tabular}
	\end{tabular}
		\caption{Covariance matrices resulting from DNS of turbulent channel flow with $Re=186$ (left plots); and the solution to optimization problem~\ref{eq.CP1} with $\gamma = 300$ (right plots). (a, b) Streamwise $V_{uu}$ and (c,d) streamwise/wall-normal $V_{uv}$ covariance matrices for $\bk = (2.5, 7)$. The one-point correlation profiles that are used as problem data in~\ref{eq.CP1} are marked by black lines along the main diagonals.}
	\label{fig.covariance_DNS_NM}
\end{figure}

In addition to matching available one-point correlations, we next demonstrate that the solution to optimization problem~\ref{eq.CP1} also provides good recovery of two-point correlations. These are not used as problem data in~\ref{eq.CP1} and correspond to off-diagonal entries in {\bf Figure~\ref{fig.output_covariance}}. The premultiplied energy spectrum in channel flow with $Re = 186$ peaks at $\bk = (2.5,7)$; e.g., see~\cite[Figure~12(a)]{moajovJFM12}. The left column in {\bf Figure~\ref{fig.covariance_DNS_NM}} displays the streamwise $V_{uu}$ and the streamwise/wall-normal $V_{uv}$ covariance matrices resulting from DNS at these flow conditions. The right column in {\bf Figure~\ref{fig.covariance_DNS_NM}} shows the same covariance matrices that are obtained from the solution to problem~\ref{eq.CP1}; see~\cite{zarjovgeoJFM17} for a detailed examination of wall-normal and spanwise covariance matrices. The quality of recovery depends on the choice of the regularization parameter $\gamma$ and, for $\gamma=300$, approximately $60\%$ of the DNS-generated covariance matrix $V_{\mathrm{dns}}$ can be recovered based on a relative Frobenius norm measure, $\norm{V - V_{\mathrm{dns}}}_F / \norm{V_{\mathrm{dns}}}_F$. Here, $V= CXC^*$ represents the two-point correlation matrix of velocity fluctuations resulting from problem~\ref{eq.CP1}. The high-quality recovery of two-point correlations is attributed to the structural constraint in Equation~\ref{eq.constraint_lyap}, which keeps physics in the mix and enforces consistency between data and the linearized NS dynamics.
	\begin{marginnote}[]
While the diagonal entries of $V$ determine the kinetic energy and affect the mean momentum transfer in the turbulent flow, the off-diagonal two-point correlations are indicators of coherent flow structures that reside at various locations away from the wall~\cite{monstewilcho07,smimckmar11}.
	\end{marginnote}

	\vspace*{-4ex}
\subsection{Stochastic linear simulations}
	\label{sec.stochastic_sim}

\begin{figure}
\centering
	\begin{tabular}{crccrc}
	\subfigure[]{\label{fig.rankZ_gamma1e4}}
	&&&
	\subfigure[]{\label{fig.stochastic_sim}}
	&&
	\\[-.4cm]
	&
	\hspace{-.85cm}
	\begin{tabular}{c}
		\vspace{.4cm}
		{\normalsize \rotatebox{90}{$\sigma_i(Z)$}}
	\end{tabular}
	&
	\hspace{-4.3cm}
	\begin{tabular}{c}
		\includegraphics[width=6cm]{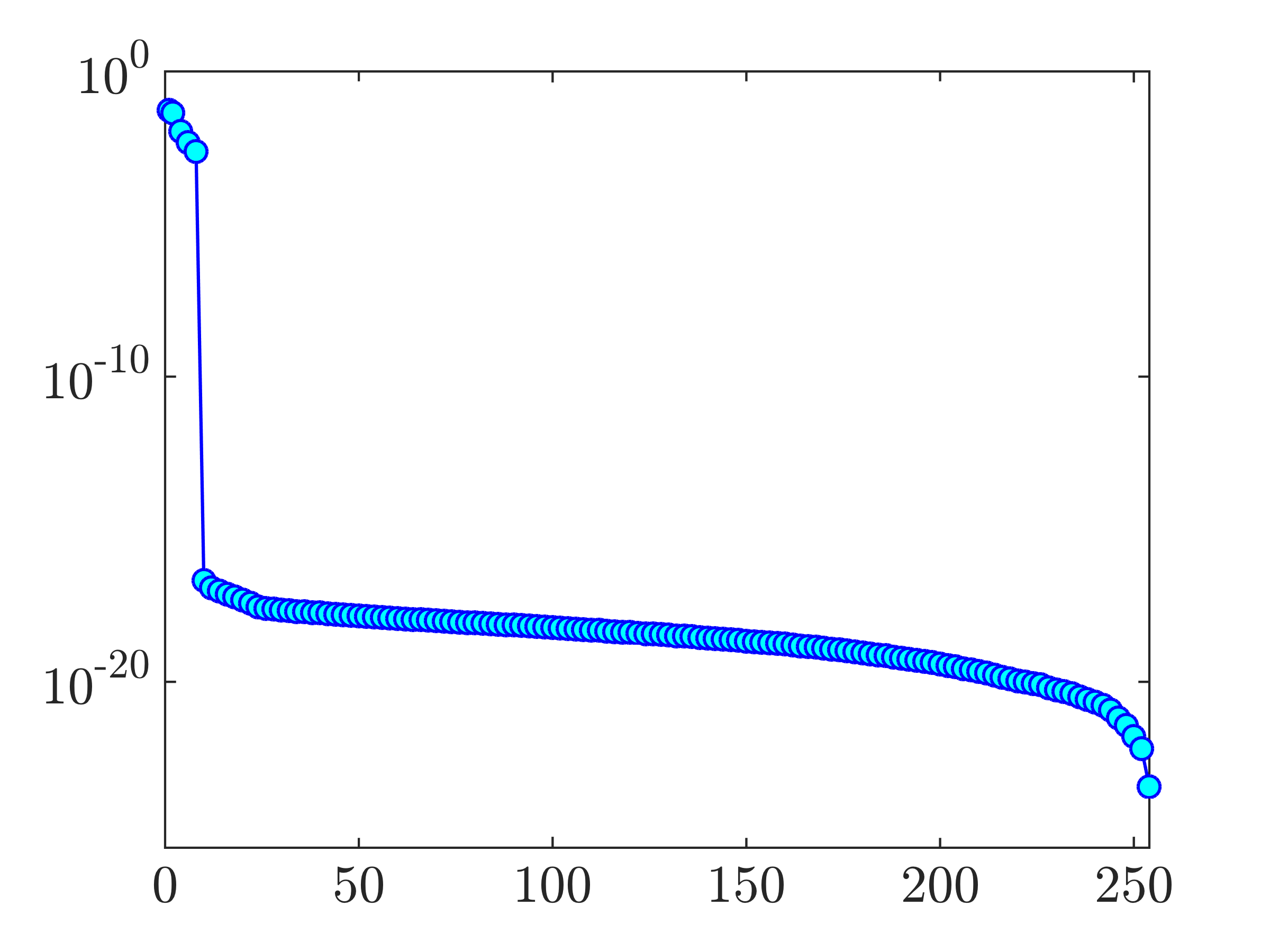}
		\\[-.1cm]
		{\normalsize $i$}
	\end{tabular}
	\hspace{-4.3cm}
	&&
	\hspace{-.85cm}
	\begin{tabular}{c}
		\vspace{.6cm}
		{\normalsize \rotatebox{90}{kinetic energy}}
	\end{tabular}
	&
	\hspace{-4.3cm}
	\begin{tabular}{c}
		\includegraphics[width=6cm]{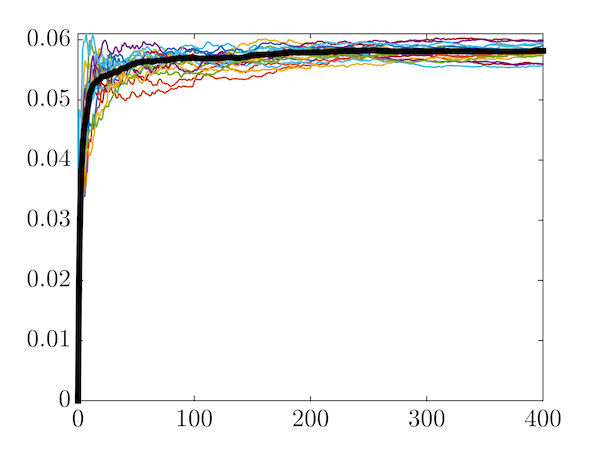}
		\\[-.1cm]
		{\normalsize $t$}
	\end{tabular}
	\end{tabular}
\caption{(a) Singular values of the solution $Z$ to~\ref{eq.CP1} in turbulent channel flow with $Re = 186$, $\bk = (2.5, 7)$, and $N = 127$ for $\gamma=10^4$. (b) Time evolution of fluctuation's kinetic energy for twenty realizations of the stochastic input to the resulting modified linearized dynamics in Equation~\ref{eq.feedback_dyn}; the energy averaged over all simulations is marked by the thick black line.}
\label{fig.svdZ_kx2p5_kz7-stochastic_sim}
\end{figure}

Stochastic simulations of the modified LTI dynamics in Equation~\ref{eq.feedback_dyn} can be used to verify the theoretical predictions resulting from the modeling and optimization framework of Section~\ref{sec.cc-noisemodeling}. For a spatial discretization with $N=127$ collocation points in the wall-normal direction, at each wavenumber $\bk$, the LTI system in Equation~\ref{eq.feedback_dyn} has $254$ states. For $\bk = (2.5, 7)$ and $\gamma = 10^4$, the matrix $Z$ that solves optimization problem~\ref{eq.CP1} has $8
$ non-zero eigenvalues ($6$ positive and $2$ negative); see {\bf Figure~\ref{fig.rankZ_gamma1e4}}. As shown in~\cite{zarchejovgeoTAC17}, the maximum number of positive or negative eigenvalues of the matrix $Z$ bounds the number of inputs into the linearized NS model given by Equation~\ref{eq.lnse1}. This implies that partially available statistics can be reproduced with $6$ colored-in-time inputs and as a result, the dynamical modification $BK$ in Equation~\ref{eq.feedback_dyn} is of rank~$6$. 

\begin{figure}
\begin{center}
	\begin{tabular}{crccrc}
		\subfigure[]{\label{fig.uu_sim}} 
		&&&
		\subfigure[]{\label{fig.vv_sim}} 
		&&
		\\[-.5cm]
		&
		\hspace{-.9cm}
		\begin{tabular}{c}
			\vspace{.45cm}
			{\normalsize \rotatebox{90}{$\diag \left( V_{uu} (\bk) \right)$}}
		\end{tabular}
		&
		\hspace{-4.4cm}
		\begin{tabular}{c}
		        \includegraphics[width=6cm]{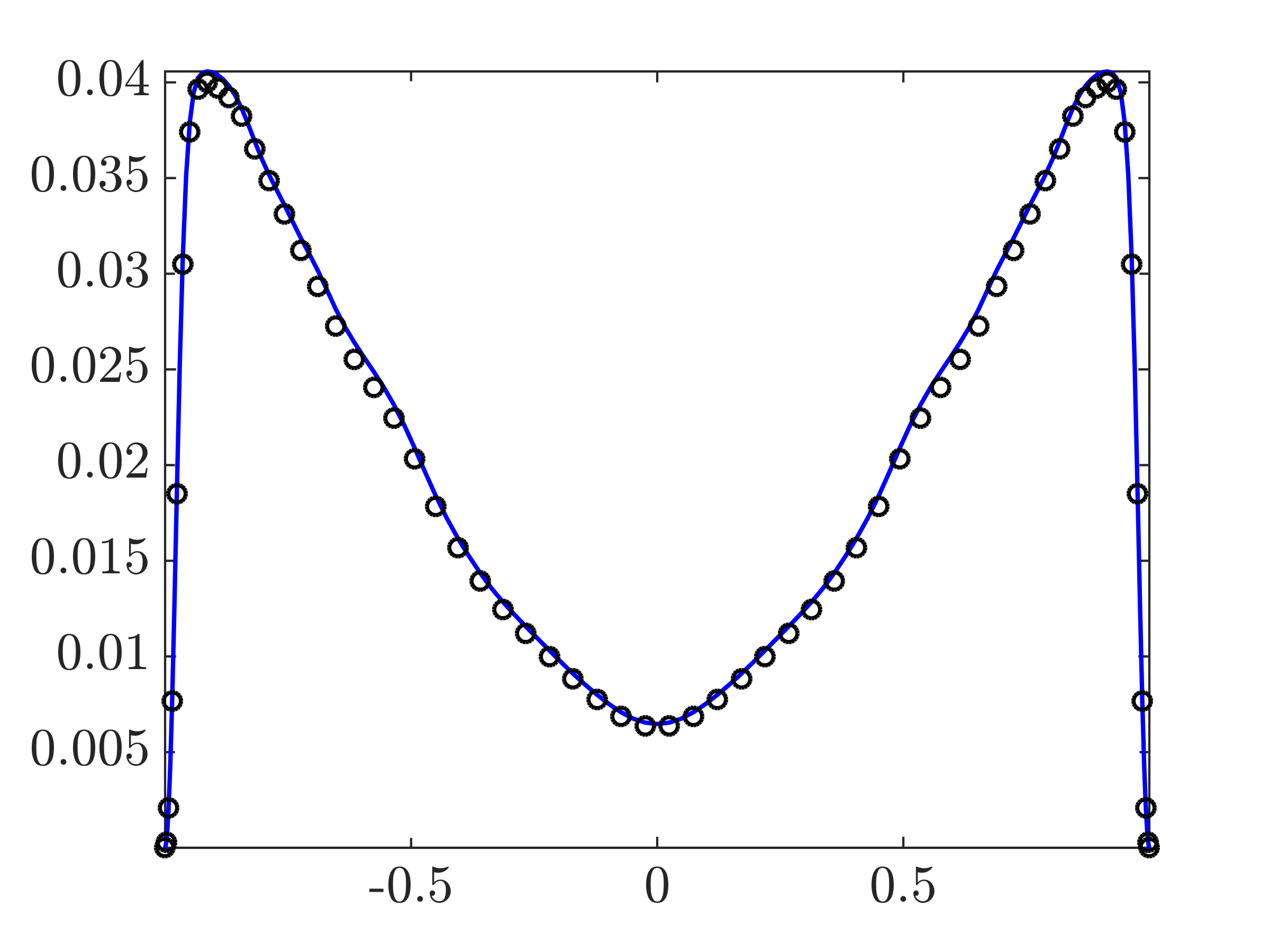}
	        	\\[-.2cm]
		\hspace{.1cm}
		        {\normalsize$y$}
	        \end{tabular}
	        \hspace{-4.2cm}
		&&
		\hspace{-.9cm}
	        \begin{tabular}{c}
	        	\vspace{.4cm}
	        	{\normalsize \rotatebox{90}{$\diag \left( V_{vv} (\bk) \right)$}}
		\end{tabular}
		&
		\hspace{-4.45cm}
	        \begin{tabular}{c}
	        	\includegraphics[width=6cm]{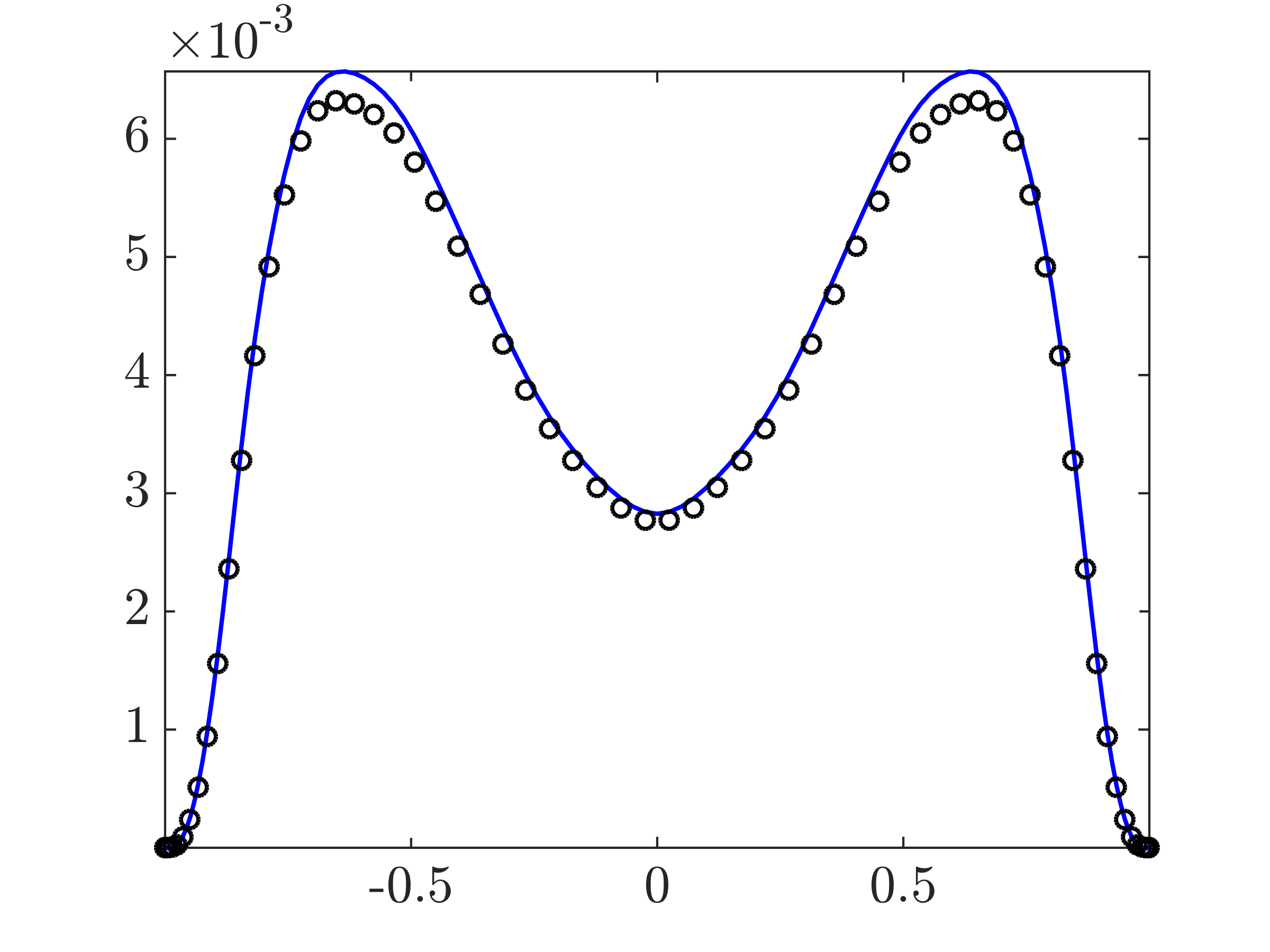}
			\\[-.2cm]
			\hspace{.1cm}
	        	{\normalsize $y$}
	        \end{tabular}
	        \\[-.2cm]
	        \subfigure[]{\label{fig.ww_sim}} 
		&&&
		\subfigure[]{\label{fig.uv_sim}} 
		&&
		\\[-.5cm]
		&
		\hspace{-.9cm}
		\begin{tabular}{c}
			\vspace{.45cm}
			{\normalsize \rotatebox{90}{$\diag \left( V_{ww} (\bk) \right)$}}
		\end{tabular}
		&
		\hspace{-4.4cm}
		\begin{tabular}{c}
		        \includegraphics[width=6cm]{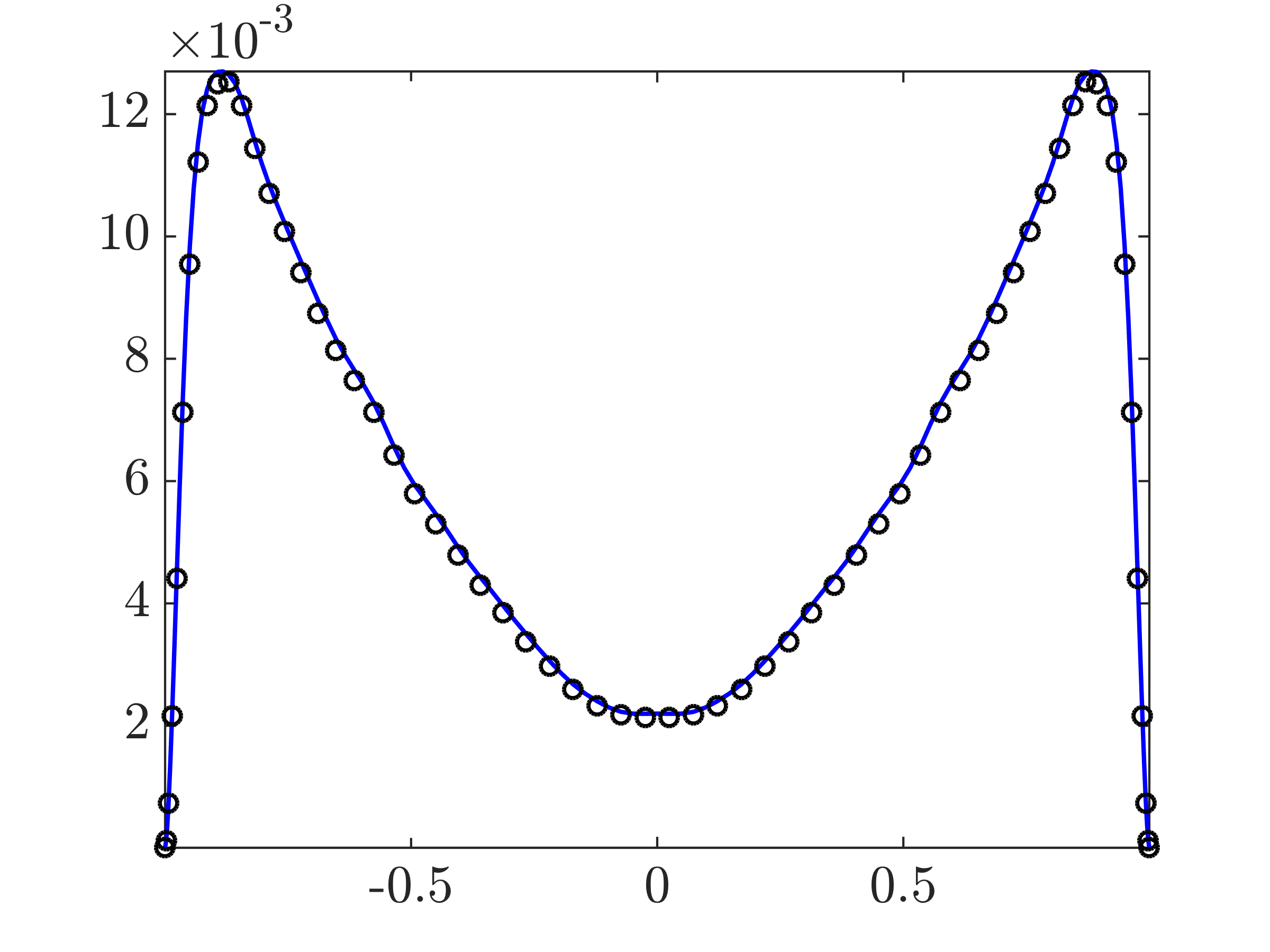}
	        	\\[-.2cm]
		\hspace{.1cm}
		        {\normalsize$y$}
	        \end{tabular}
	        \hspace{-4.2cm}
		&&
		\hspace{-.9cm}
	        \begin{tabular}{c}
	        	\vspace{.4cm}
	        	{\normalsize \rotatebox{90}{$\diag \left( V_{uv} (\bk) \right)$}}
		\end{tabular}
		&
		\hspace{-4.45cm}
	        \begin{tabular}{c}
	        	\includegraphics[width=6cm]{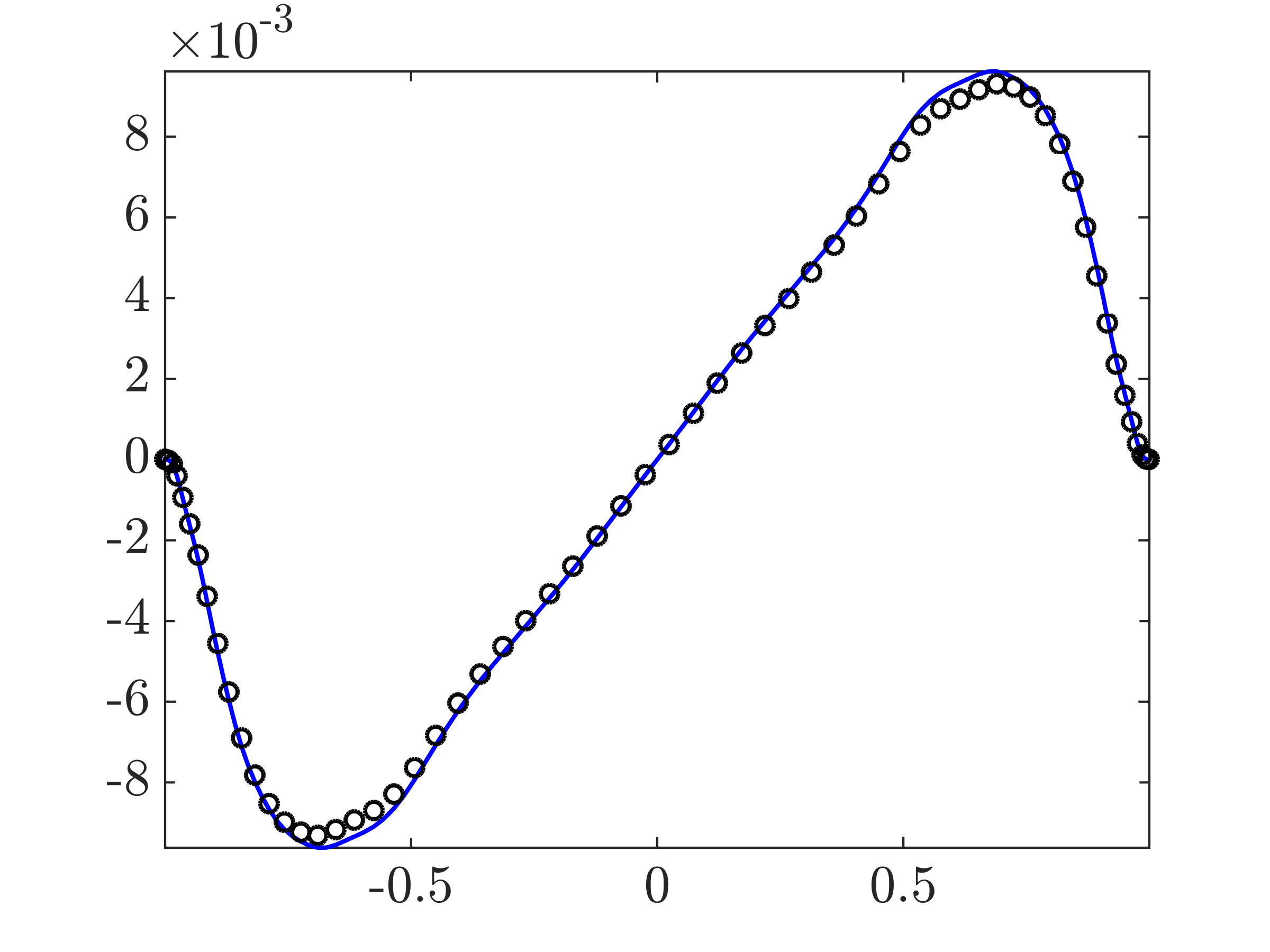}
			\\[-.2cm]
			\hspace{.1cm}
	        	{\normalsize $y$}
	        \end{tabular}
        \end{tabular}
\end{center}
\caption{Normal stress profiles in the (a) streamwise, (b) wall-normal, and (c) spanwise directions; (d) shear stress profile resulting from DNS of turbulent channel flow with $Re=186$ at $\bk=(2.5,7)$ (--) and stochastic linear simulations ($\Circle$).}
\label{fig.sim_results}
\end{figure}

Proper comparison with DNS or experiments requires ensemble-averaging, rather than comparison at the level of individual stochastic simulations. To this end, twenty simulations with different realizations of white-in-time input $\bw (\bk,t)$ in Equation~\ref{eq.feedback_dyn} have been conducted. The total simulation time was $400$ viscous time units. {\bf Figure~\ref{fig.stochastic_sim}} shows the time evolution of the energy (variance) of velocity fluctuations resulting from these twenty simulations. The variance averaged over all simulations is marked by the thick black line. Even though the responses of individual simulations differ from each other, the average of twenty sample sets asymptotically approaches the correct value of turbulent kinetic energy in statistical steady-state, $\trace\,  (V (\bk))$. {\bf Figure~\ref{fig.sim_results}} displays the normal and shear stress profiles resulting from DNS and from stochastic linear simulations. The averaged output of the twenty simulations agrees well with DNS results. This agreement can be further improved by running additional simulations and by increasing the total simulation times.

\begin{figure}
\begin{center}
	\begin{tabular}{crccrc}
		\subfigure[]{\label{fig.FreqResp_PSD_ArtAedAf_R186_kx2p5_kz7}}
		&&&
		\subfigure[]{\label{fig.FreqResp_maxsig_ArtAedAf_R186_kx2p5_kz7}}
		&&
		\\[-.5cm]
		&
		\hspace{-.8cm}
		\begin{tabular}{c}
			\vspace{.45cm}
			{\normalsize \rotatebox{90}{$\Pi_\bv (\bk,\omega)$}}
		\end{tabular}
		&
		\hspace{-4.4cm}
		\begin{tabular}{c}
		        \includegraphics[width=5.8cm]{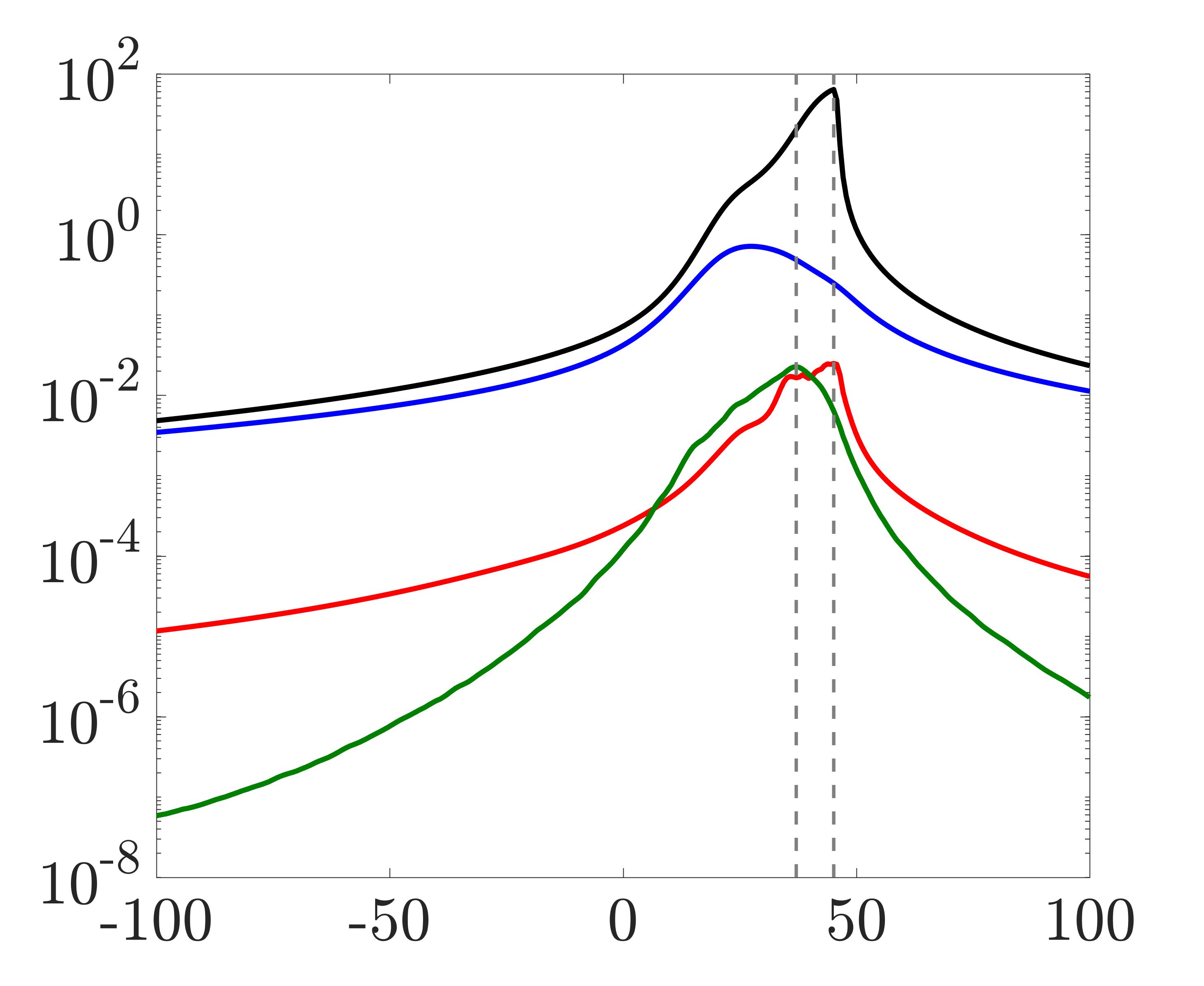}
	        	\\[-.2cm]
		\hspace{.1cm}
		        {\normalsize$\omega$}
	        \end{tabular}
	        \hspace{-4.2cm}
		&&
		\hspace{-.8cm}
	        \begin{tabular}{c}
	        	\vspace{.4cm}
	        	{\normalsize \rotatebox{90}{$y^+$}}
		\end{tabular}
		&
		\hspace{-4.55cm}
	        \begin{tabular}{c}
	        	\includegraphics[width=5.8cm]{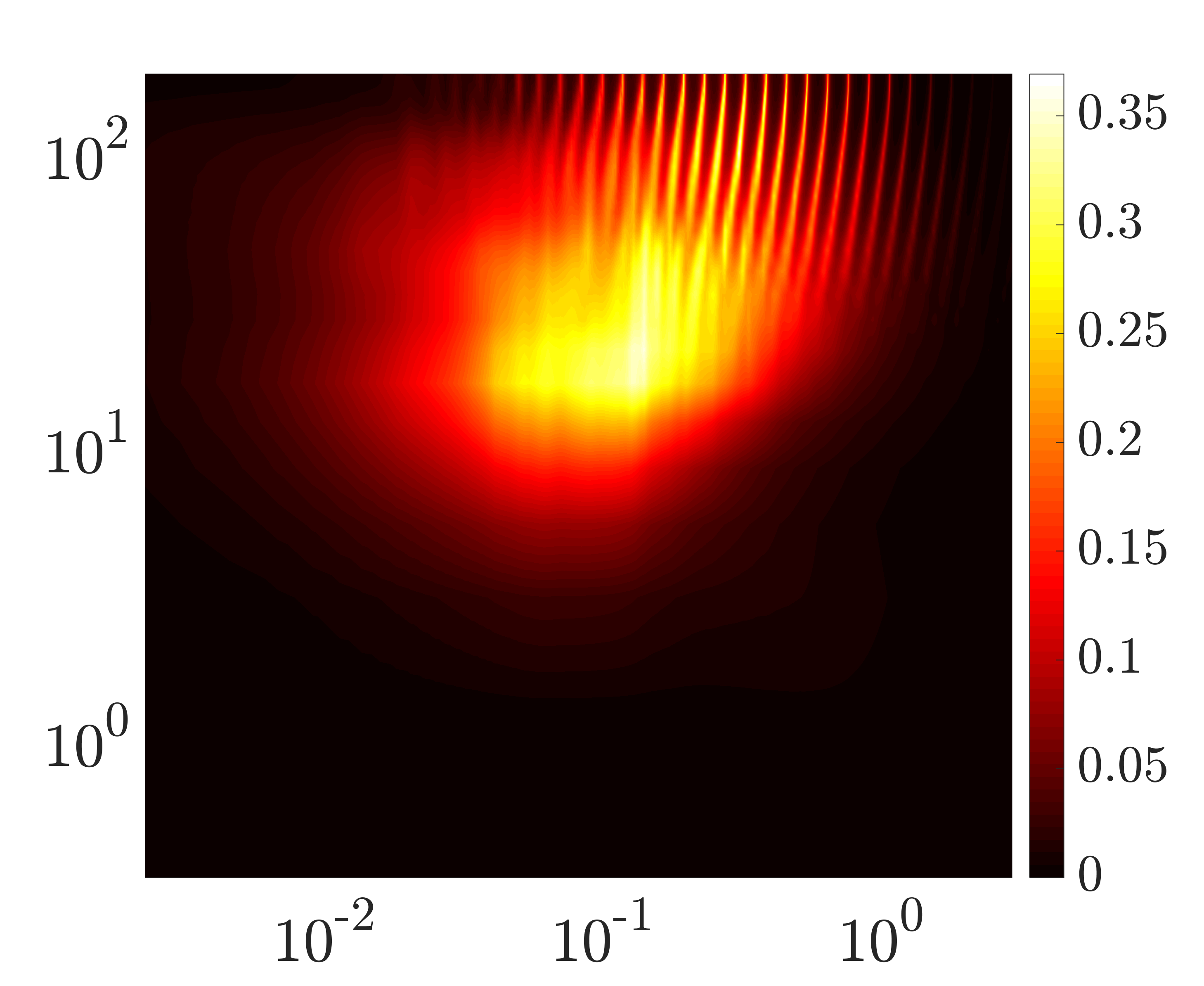}
			\\[-.2cm]
			\hspace{.1cm}
	        	{\normalsize $\omega^+$}
	        \end{tabular}
	\end{tabular}
\end{center}
\caption{(a) Power spectral density $\Pi_\bv (\bk,\omega)$ resulting from DNS of turbulent channel flow with $Re =186$ at $\bk=(2.5,7)$ (green), the linearized NS model in Equation~\ref{eq.lnse1} (black), an eddy-viscosity-enhanced linearized model (blue), and the modified LTI dynamics in Equation~\ref{eq.feedback_dyn} for $\gamma=300$ (red). (b) Premultiplied energy spectrum of the modified dynamics in Equation~\ref{eq.feedback_dyn} for turbulent channel flow with $Re=186$ resulting from the integration of $\omega \, \diag \, ( T_{\bv \bw} (\bk, \omega) T^*_{\bv\bw} (\bk, \omega))$ over wavenumbers $\bk$.}
\label{fig.FreqResp_ArtAedAf}
\end{figure}

	\vspace*{-2ex}
\subsection{Spatio-temporal energy spectrum}
	\label{sec.frequency-resp}

To analyze the spatio-temporal aspect of dynamical models resulting from the framework of Section~\ref{sec.cc-noisemodeling} we examine the Power Spectral Density (PSD) and the energy spectrum of velocity fluctuations. The PSD of the LTI system in Equation~\ref{eq.feedback_dyn} is determined by the sum of squares of the singular values of the frequency response matrix in Equation~\ref{eq.Tvw},
\begin{align*}
	\Pi_\bv (\bk,\omega)
	\;=\;
	\trace \left( T_{\bv \bw} (\bk, \omega)\, T_{\bv \bw}^* (\bk, \omega) \right)
	\; = \;
	\sum_i
	\, 
	\sigma_i^2 (T_{\bv \bw} (\bk, \omega)).
\end{align*}
Integration of $\Pi_\bv (\bk,\omega)$ over the temporal frequencies yields the square of the $H_2$ norm of the system in Equation~\ref{eq.feedback_dyn} or, equivalently, the $\bk$-parameterized energy spectrum~\cite{jovbamJFM05},
	\begin{align*}
	E(\bk) 
	\; = \; 
	\dfrac{1}{2} 
	\, 
	\int_{-\infty}^{\infty} \Pi_\bv(\bk,\omega) \, \mrd \omega 
	\; = \;
	\dfrac{1}{2} 
	\, 
	\trace \left( V (\bk) \right).
	\end{align*}
For a turbulent channel flow with $Re = 186$ and $\bk=(2.5,7)$, {\bf Figure~\ref{fig.FreqResp_PSD_ArtAedAf_R186_kx2p5_kz7}} compares the power spectral densities of the linearized NS model given by Equation~\ref{eq.lnse1}, the eddy-viscosity-enhanced modification of the linearized NS equations~\cite{reyhus72-3,deljimjfm06,cospujdep09,pujgarcosdep09,hwacosJFM10b}, and the dynamical model given by Equation~\ref{eq.feedback_dyn} resulting from the framework presented in Section~\ref{sec.cc-noisemodeling} with the result of DNS. For the first two models, the input matrix $B(\bk)$ excites all degrees of freedom in the state equation and, for the modified dynamics, the input matrix $B(\bk)$ comes from the framework presented in Section~\ref{sec.cc-noisemodeling} with the regularization parameter $\gamma = 300$. All three models are driven by spatially and temporally uncorrelated inputs. 

The temporal frequency at which the PSD peaks is similar for the linearized NS equations and the modified dynamics ($\omega \approx 45$) and is closer to the result of DNS ($\omega \approx 37$) than the frequency associated with the eddy-viscosity-enhanced model ($\omega \approx 27$). We also see that both the eddy-viscosity enhancement and the data-driven low-rank modification attenuate the amplification of disturbances at all temporal frequencies. The uniform damping of the PSD ensures that the $H_2$ norm of the system in Equation~\ref{eq.feedback_dyn} matches the energy spectrum of the turbulent channel flow; cf.\ red and green curves in {\bf Figure~\ref{fig.FreqResp_PSD_ArtAedAf_R186_kx2p5_kz7}}. For the modified dynamics given by Equation~\ref{eq.feedback_dyn}, {\bf Figure~\ref{fig.FreqResp_maxsig_ArtAedAf_R186_kx2p5_kz7}} shows the premultiplied spatio-temporal energy spectrum as a function of the wall-normal coordinate and temporal frequency in inner (viscous) units, i.e., $y^+ \DefinedAs (1+y) Re$ and $\omega^+ \DefinedAs \omega/Re$. This spectrum is computed by integrating $\omega \, \diag \, ( T_{\bv \bw} (\bk, \omega) T^*_{\bv\bw} (\bk, \omega))$ over $\bk$ and is concentrated around $y^+ \approx 15$ within a frequency band $\omega^+ \in (0.01, 1)$, which is in agreement with the trends observed in DNS-generated energy spectra~\cite{towlozyan19}. Improving the accuracy in matching the temporal correlations resulting from DNS may require closer examination of the role of parameter $\gamma$ or the addition of extra constraints in problem~\ref{eq.CP1} and is a subject of ongoing research.

%%\newpage
% Summary Points
	\vspace*{-1ex}
\begin{summary}[SUMMARY POINTS]
\begin{enumerate}
\item Data from numerical simulations and experiments can be used to refine the predictive power of models that arise from first principles, e.g., the linearized NS equations. 

\item White-in-time stochastic input to the linearized NS equations cannot explain second-order statistics of turbulent wall-bounded flows.

\item Colored-in-time stochastic input that excites all degrees of freedom can completely cancel the original dynamics and yield a model that does not generalize well. 

\item A suitably regularized solution to covariance completion problems can ensure that important features of spatio-temporal responses are captured via low-complexity stochastic dynamical models.

\item The effect of colored-in-time stochastic input can be equivalently interpreted as a structural perturbation of the linearized dynamical generator, which can be used to identify important state-feedback interactions that are lost through linearization.

\item Combining tools and ideas from systems theory and convex optimization can pave the way for the systematic development of theory and techniques that combine data-driven with physics-based modeling. 
\end{enumerate}
\vspace*{-2ex}
\end{summary}

% Future Issues
	\vspace*{-4ex}
\begin{issues}[FUTURE ISSUES]
\begin{enumerate}
\item Modeling of flow disturbances plays an important role in obtaining well-possed estimation gains~\cite{hoechebewhen05,chehoebewhen06}. Stochastic dynamical models that are obtained via covariance completion fit nicely into a Kalman filtering framework for turbulent flows and have the potential to open the door for a successful output-feedback design at higher Reynolds numbers than current feedback~\citep{bewliu98,hogbewhen03,kimbew07} and sensor-free~\citep{fratalbracos06,jovPOF08,moajovJFM10,liemoajovJFM10,moajovJFM12} strategies allow. The efficacy of such an approach and its interplay with real-time estimation and feedback control are yet to be examined.

\item Turbulence modeling for complex fluids and flows in complex geometries~\cite{hodjovkumJFM08,hodjovkumJFM09,jovkumPOF10,jovkumJNNFM11,liejovkumJFM13,jeunicjovPOF16,hildwinicjovcanPRF18,dwisidniccanjovJFM19} requires dealing with a large number of degrees of freedom. Since improving upon current algorithms that require $O(n^3)$ computations for a model with $n$ states is challenging, a possible direction is to examine physical approximations~\cite{reesararn96,her97,hoghen02,ranzarhacjovPRF19a} and model reduction techniques~\cite{row05,sch10,jovschnicPOF14,rowdaw17}.

\item The regularization terms in problems~\ref{eq.CP1} and~\ref{eq.CP2} are used as convex surrogates for rank and cardinality. For problems with structural constraints such surrogates do not enjoy standard probabilistic guarantees~\cite{canrec09}, and the utility of more refined approximations techniques, e.g., manifold optimization~\cite{absmahsep08}, low-rank inducing norms~\cite{gruzarjovranCDC16,grurangis18}, and nonconvex matrix completion~\cite{canlisol15,sunluo16,geleema16} in low-complexity stochastic dynamical modeling remains largely unexplored.

\item Higher-order turbulent flow statistics often play an important role in characterizing quantities of interest in engineering applications; e.g., fourth-order statistics are relevant in acoustic source modeling for high-speed jets~\cite{karafshyndowmcmpokpagmcg10,leigol11}. The importance of matching higher-order statistics calls for a generalized theory for the stochastic realization of state-statistics that are currently limited to second-order correlations.

\item The output of the stochastically-forced linear model can be used to drive the mean flow equations in time-dependent stochastic simulations. It is important to identify conditions under which the feedback interconnection in {\bf Figure~\ref{fig.equilibrium2}} converges.
\end{enumerate}
\vspace*{-2ex}
\end{issues}

%Conclusions
	 \vspace*{-6ex}
\section{CONCLUDING REMARKS}
\label{sec.conclusions}

This review discusses a framework that combines tools from systems theory and optimization to develop low-complexity models of turbulent flows that are well-suited for analysis and control synthesis. The goal is to embed partially known statistical signatures obtained via numerical simulation of the NS equations or experimental measurements into first principles models that arise from linearization around the turbulent mean velocity. This amounts to identifying the spectral content of stochastic excitation into the linearized equations such that turbulent statistics can be reproduced. The review focuses on the completion of second-order statistics and while the methodology and theoretical framework are applicable to a wide range of scenarios, a channel flow configuration is used to solidify the discussion. On par with the dramatic upswing from the fields of machine learning and optimization in leveraging big-data for modeling, the proposed methodology utilizes data to refine the predictive capability of a dynamical model that arises from first principles and it offers a new perspective on tackling issues of robustness and generalizability. 

%Disclosure
	\vspace*{-3ex}
\section*{DISCLOSURE STATEMENT}
The authors are not aware of any biases that might be perceived as affecting the objectivity of this review.

% Acknowledgements
\vspace*{-3ex}
\section*{ACKNOWLEDGMENTS}
Financial support from NSF under Awards CMMI 1739243 and ECCS 1809833, and AFOSR under Awards FA9550-16-1-0009 and FA9550-18-1-0422 is gratefully acknowledged. We thank Anubhav Dwivedi for generating DNS results reported in Section~\ref{sec.frequency-resp}.

% References

\end{document}